# An efficient topology optimization method based on adaptive reanalysis with projection reduction

Jichao Yin[*], Hu Wang[†], Shuhao Li, Daozhen Guo

*State Key Laboratory of Advanced Design and Manufacturing for Vehicle Body, Hunan University, Changsha, 410082, P.R. China*

**Abstract**

An efficient topology optimization based on the adaptive auxiliary reduced model reanalysis (AARMR) method is proposed to improve computational efficiency and scale. In this method, a projection auxiliary reduced model (PARM) is integrated into the combined approximation reduced model (CARM) to reduce the dimension of the model in different aspects. First, the CARM restricts the solution space to avoid large matrix factorization. Second, the PARM is proposed to construct the CARM dynamically to save computational cost. Furthermore, the multi-grid conjugate gradient method is suggested to update PARM adaptively. Finally, several classic numerical examples are tested to show that the proposed method not only significantly improves computational efficiency, but also can solve large-scale problems that are difficult to solve by direct solvers due to the memory limitations.

---

[*] First author. E-mail address: jcyin@hnu.edu.cn (J. Yin)

[†] Corresponding author. Tel: +86 0731 88655012; fax: +86 071 88822051,
 E-mail address: wanghu@hnu.edu.cn (H. Wang)

# 1. Introduction

Topology optimization is widely applied in several engineering fields to provide conceptual designs without experience, ensuring the best trade-off between stiffness and weight by fully exploiting material properties. One of the primary challenges in engineering applications is the computational burden when encountering large-scale problems[1]. The computational cost of each evaluation of topology optimization is mainly governed by three aspects: modeling, analysis (repeated calculation), and updating design variables[2]. With the increase in degrees of freedom (DOFs), analysis has become the dominant computational burden, especially for 3D problems with more than millions of DOFs[3]. It is prohibitive to handle large-scale problems by direct solvers. Therefore, researchers are motivated to search for a way to alleviate the computational burden.

Iterative methods and reduced-order models (ROMs) may be the most attractive issues, such as the preconditioned Krylov subspace method[4], multigrid preconditioned conjugate gradients (MGCG)[5], combinatorial approximate (CA) reanalysis[3], and on-the-fly ROM[6]. However, the condition number of the stiffness matrix is relatively large during topology optimization, which will cause iterative solvers to converge slowly. The ROM is employed in this study because it only performs matrix multiplication when solving equilibrium equations. Approximate reanalysis[7] is an ROM suitable for structural topology modification, which is compatible with iterative redesign in topology optimization. Kirsch[8] pointed out that combined approximation can achieve excellent results with small computational effort for large changes in the cross-sections and structural topology. Huang[9] et al. proposed a multi-grid assisted reanalysis method (MGR) to perform structural response analysis when the grid is modified. In recent years, the CA method proposed by Kirsch[10] for linear static reanalysis has become one of the most concerning reanalysis methods. The CA[11] integrates the accuracy of the global approximation[12] and the efficiency of the local approximation[13]; more details can be found in literatures [8, 14]. Kirsch[15] et al. proved that the results obtained by the CA are equivalent to the preconditioned conjugate gradient (PCG) method. To date, CA has been successfully applied in dynamic[16], nonlinear[17], and other fields[18-20].

Moreover, Amir[21] et al. utilized CA to improve the computational efficiency of robust topology optimization without affecting the optimization process results. Bogomolny[22] applied CA to save the computational cost of repeated eigenvalue analysis involved in free vibration topology optimization. The block combined approximation with shifting (BCAS) method was introduced by Zheng[23] et al. to handle topology optimization under multi-frequency harmonic force excitations. Long[24] et al. proposed a high-efficiency approximate reanalysis method based on the introduction of reciprocal-type variables to solve the topology optimization subject to multiple constraints. Mo[25] proposed an iterative reanalysis approximation for moving morphable components (MMCs) to reduce evaluations of topology optimization redesign. Although the CA has been successfully applied to the topology optimization field, there are still bottlenecks in solving large-scale problems to the best of our knowledge. The main issues are summarized below.

1) The reanalysis approximation process cannot accommodate extremely large changes in stiffness matrix. However, the evolution of the density field during topology optimization means that the stiffness changes greatly, especially in the early iterations. In other words, the approximate solutions obtained by the combinatorial approximation usually does not meet the optimization requirements after several iterations, and the stiffness matrix needs to be re-decomposed and the combined approximation reduced model (hereinafter referred to as CARM) is reconstructed.

2) The computational cost of large matrix factorization and other operations involved in constructing CARM is very expensive, and the hardware requirements for storing the inverse matrix are burdensome.

Therefore, this study proposed an efficient topology optimization method with adaptive auxiliary reduced model reanalysis (AARMR), which overcomes the above issues simultaneously. The AARMR method employs two reduced models to achieve the dimension reduction of structural models in different aspects. A projection auxiliary reduced model (hereinafter referred to as PARM) is used to avoid complex operations for matrix decomposition and others. In this way, the computational cost is significantly reduced, and the hardware burden is alleviated. The CARM is used to restrict the solution space to obtain approximate solutions and is constructed dynamically at each iteration. The cost of construction of CARM is relatively cheap due to the introduction of PARM. Moreover, the distortion caused by a large number of DOF changes can be naturally overcome. This paper is organized as follows. Standard topology optimization is briefly reviewed in Section 2 and the proposed adaptive auxiliary reduced model reanalysis method is described in detail. Section 3 integrates the proposed method into the topology optimization framework. Several classic examples are implemented to verify the efficiency and accuracy of the proposed method in Section 4. Finally, some conclusions are given in Section 5.

## 2. Overview of theoretical formulation

The primary purpose is of this study to improve the solution efficiency of topology optimization, especially for large-scale 3D problems. For this purpose, the adaptive auxiliary reduced model reanalysis (AARMR) method is proposed and integrated into the topology optimization framework.

### 2.1 Topology optimization formulation

The concept of topology optimization was originally introduced by Bendsøe and Kikuchi[26]. In recent decades, a variety of topology optimization approximations have been proposed. In this study, the modified solid isotropic material interpolation with penalization (SIMP) method[27] is employed.

$$E_e = E_{\min} + (E_0 - E_{\min})\tilde{\rho}_e^p, \qquad (1)$$

where $E_0$ is the Young's modulus of the solid elements. Weak elements with Young's modulus $E_{\min}$ are used to avoid ill-conditioned problems ($E_0 \gg E_{\min}$). $p$ is the penalty factor. In addition, filter techniques are used to suppress checkerboard phenomena. The physical variables $\tilde{\rho}$ used in the optimization process are obtained by filtering the design variables $\boldsymbol{\rho}$.

$$\tilde{\rho}_e = f(\rho_i), \quad \rho_i \in \Omega_e, \tag{2}$$

where $f(\cdot)$ represents a filter method[28-30] and $\Omega_e$ is the filter area centered on element $e$. Based on the above conditions, the topology optimization formula for minimizing end-compliance can be expressed as:

$$\begin{aligned}
&\min: C = \boldsymbol{u}^T \mathbf{K}(\tilde{\boldsymbol{\rho}})\boldsymbol{u}, \\
&s.t. : \mathbf{K}(\tilde{\boldsymbol{\rho}})\boldsymbol{u} = \boldsymbol{f}, \\
&\quad\quad \sum v_e \tilde{\rho}_e - V \leq 0, \\
&\quad\quad 0 \leq \rho_e \leq 1, \quad e = 1, 2, ..., N,
\end{aligned} \tag{3}$$

where $\mathbf{K}$ is the global stiffness matrix, $\boldsymbol{u}$ and $\boldsymbol{f}$ are the displacement and load vectors, respectively. $v_e$ is the element volume, and $V$ is an allowed volume fraction determined by the users. The optimization problem is solved by sequential approximation approaches such as OC[31, 32] and MMA[33] to update the design variables. The derivative of the objective function with respect to design variables is the key to sequential approximation approaches. Calculating the derivative by using the chain rule can be written as:

$$\frac{\partial C}{\partial \rho_e} = \sum_{i \in N_e} \frac{\partial(\boldsymbol{u}^T \mathbf{K} \boldsymbol{u})}{\partial \tilde{\rho}_i} \frac{\partial \tilde{\rho}_i}{\partial \rho_e} = \sum_{i \in N_e} (2\boldsymbol{u}^T \mathbf{K} \frac{\partial \boldsymbol{u}}{\partial \tilde{\rho}_i} + \boldsymbol{u}^T \frac{\partial \mathbf{K}}{\partial \tilde{\rho}_i} \boldsymbol{u}) \frac{\partial \tilde{\rho}_i}{\partial \rho_e}. \tag{4}$$

It should be noted that $\boldsymbol{f}$ is assumed to be design-independent. The derivative of the displacement vector with respect to the physical variable can be obtained by deriving the equilibrium equation,

$$\frac{\partial \mathbf{K}}{\partial \tilde{\rho}_i} \boldsymbol{u} + \mathbf{K} \frac{\partial \boldsymbol{u}}{\partial \tilde{\rho}_i} = \boldsymbol{0}. \tag{5}$$

Substituting Equation (5) into Equation (4), the derivative can be obtained as follows:

$$\frac{\partial C}{\partial \rho_e} = \sum_{i \in N_e} (-\boldsymbol{u}^T \frac{\partial \mathbf{K}}{\partial \tilde{\rho}_i} \boldsymbol{u}) \frac{\partial \tilde{\rho}_i}{\partial \rho_e}. \tag{6}$$

## 2.2 Reanalysis by adaptive auxiliary reduced model

Equation (7) which characterizes the structure is called repeatedly as a function.

Materials within the design domain are redistributed in each iteration, resulting in the modified global stiffness matrix,

$$\mathbf{K}u = f. \tag{7}$$

From introducing the change in stiffness matrix $\Delta \mathbf{K} = \mathbf{K} - \mathbf{K}_0$, it follows that Equation (7) may be rewritten for the redesigned structure as:

$$(\mathbf{K}_0 + \Delta \mathbf{K})u = f, \tag{8}$$

and hence

$$\mathbf{K}_0 u = f - \Delta \mathbf{K} u. \tag{9}$$

From Equation (9), the recurrence relation $\mathbf{K}_0 u^{(k)} = f - \Delta \mathbf{K} u^{(k-1)}$ is defined, and the displacement vector can be written as:

$$u^{(k)} = u_0 - \mathbf{K}_0^{-1} \Delta \mathbf{K} u^{(k-1)}. \tag{10}$$

Here, $\mathbf{K}_0$ represents the reference stiffness matrix, and $u_0$ is reference solution obtained by solving the equilibrium equation $\mathbf{K}_0 u_0 = f$. Therefore, the displacement vector can be approximated based on the so-called binomial series expansion as:

$$\begin{aligned} u &= u_0 + (-\mathbf{K}_0^{-1} \Delta \mathbf{K}) u_0 + (-\mathbf{K}_0^{-1} \Delta \mathbf{K})^2 u_0 + (-\mathbf{K}_0^{-1} \Delta \mathbf{K})^3 u_0 + ... \\ &= u_0 + \mathbf{B}^1 u_0 + \mathbf{B}^2 u_0 + \mathbf{B}^3 u_0 + ..., \end{aligned} \tag{11}$$

where

$$\mathbf{B} = -\mathbf{K}_0^{-1} \Delta \mathbf{K}. \tag{12}$$

It should be clear that the "reference stiffness matrix" refers to the stiffness matrix that has been decomposed before the new decomposition is performed. Therefore, the reference stiffness matrix is not fixed, but dynamically updated with some criterion. The main feature of CA is the utilization of the series sequence[8] as basis vectors of CARM (**R**), thus building an approximate solution of the system (7), expressed in the form of a linear combination of the series sequence. Usually, only the first $m$ items are considered, and introducing the coefficient vector $y$ as the weights of the linear combination, the solution can be approximately expressed as:

$$\begin{aligned}
\tilde{u} &= y_1 u_0 + y_2 \mathbf{B}^1 u_0 + \ldots + y_m \mathbf{B}^{m-1} u_0 \\
&= y_1 r_1 + y_2 r_2 + \ldots + y_m r_m \\
&= \mathbf{R} y,
\end{aligned} \qquad (13)$$

where $r_1 = u_0$, $r_i = \mathbf{B} r_{i-1}$ and $\mathbf{R} = [r_1, r_2, \ldots, r_m]$. Therefore, Equation (7) is rewritten as

$$\mathbf{K}\mathbf{R} y = f. \qquad (14)$$

However, the variation of design variables occurs in almost all elements at the early stage of topology optimization. This suggests that the modified stiffness matrix is relatively large. After several iterations, the accuracy of the solution may not be guaranteed, which means that the reference stiffness matrix should be replaced and the CARM should be reconstructed. Moreover, the cost of matrix decomposition and other operations of the reference stiffness matrix is prohibitive in terms of computational cost and memory requirements. It is necessary to update the reference stiffness matrix dynamically and utilize the approximate inverse instead of the exact inverse. Therefore, an auxiliary model PARM ($\mathbf{\Phi} \in \mathbb{R}^{n \times s}$, $n \gg s$) is employed to project the reference stiffness matrix into a low-dimensional matrix space,

$$\mathbf{K}_\Phi = \mathbf{\Phi}^\mathrm{T} \mathbf{K}_0 \mathbf{\Phi}. \qquad (15)$$

Taking the inverse of both sides of Equation (15)

$$\mathbf{K}_\Phi^{-1} = \mathbf{\Phi}^{-1} \mathbf{K}_0^{-1} (\mathbf{\Phi}^\mathrm{T})^{-1}. \qquad (16)$$

The pseudoinverses of PARM and its transpose are given as $\mathbf{\Phi}^{-1} = \mathbf{\Phi}^\mathrm{T}(\mathbf{\Phi}\mathbf{\Phi}^\mathrm{T})^{-1}$ and $(\mathbf{\Phi}^\mathrm{T})^{-1} = (\mathbf{\Phi}\mathbf{\Phi}^\mathrm{T})^{-1}\mathbf{\Phi}$, respectively. With these definitions, Equation (17) can be modified as:

$$\mathbf{K}_\Phi^{-1} = \mathbf{\Phi}^\mathrm{T}(\mathbf{\Phi}\mathbf{\Phi}^\mathrm{T})^{-1} \mathbf{K}_0^{-1} (\mathbf{\Phi}\mathbf{\Phi}^\mathrm{T})^{-1} \mathbf{\Phi}. \qquad (17)$$

Multiplying both sides of the equation by $\mathbf{\Phi}$ to the left and $\mathbf{\Phi}^\mathrm{T}$ to the right the approximate inverse of the reference matrix can be written as:

$$\tilde{\mathbf{K}}_0^{-1} = \mathbf{\Phi}\mathbf{\Phi}^\mathrm{T}(\mathbf{\Phi}\mathbf{\Phi}^\mathrm{T})^{-1} \mathbf{K}_0^{-1} (\mathbf{\Phi}\mathbf{\Phi}^\mathrm{T})^{-1} \mathbf{\Phi}\mathbf{\Phi}^\mathrm{T} = \mathbf{\Phi} \mathbf{K}_\Phi^{-1} \mathbf{\Phi}^\mathrm{T}. \qquad (18)$$

The dimension of matrix inversion is reduced from $\mathbb{R}^{n \times n}$ to $\mathbb{R}^{s \times s}$ after the reference stiffness matrix is projected by the PARM, and the computational cost and the memory of matrix factorization are almost negligible. Therefore, it is feasible to construct CARM dynamically, which provides a premise for applying CA to solve large-scale problems.

Substituting Equation (18) into Equation (11), the approximate solution can be expressed as:

$$\begin{aligned}\boldsymbol{u} &= \boldsymbol{u}_0 + (-\boldsymbol{\Phi}\mathbf{K}_\Phi^{-1}\boldsymbol{\Phi}^{\mathrm{T}}\Delta\mathbf{K})\boldsymbol{u}_0 + (-\boldsymbol{\Phi}\mathbf{K}_\Phi^{-1}\boldsymbol{\Phi}^{\mathrm{T}}\Delta\mathbf{K})^2\boldsymbol{u}_0 + (-\boldsymbol{\Phi}\mathbf{K}_\Phi^{-1}\boldsymbol{\Phi}^{\mathrm{T}}\Delta\mathbf{K})^3\boldsymbol{u}_0 + \ldots \\ &= (\mathbf{I} + \mathbf{C}^1 + \mathbf{C}^2 + \mathbf{C}^3 + \ldots)\boldsymbol{u}_0.\end{aligned} \quad (19)$$

The CARM can be rewritten as:

$$\begin{aligned}\mathbf{R} &= \begin{bmatrix} \boldsymbol{u}_0 & (-\boldsymbol{\Phi}\mathbf{K}_\Phi^{-1}\boldsymbol{\Phi}^{\mathrm{T}}\Delta\mathbf{K})\boldsymbol{u}_0 & (-\boldsymbol{\Phi}\mathbf{K}_\Phi^{-1}\boldsymbol{\Phi}^{\mathrm{T}}\Delta\mathbf{K})^2\boldsymbol{u}_0 \ldots & (-\boldsymbol{\Phi}\mathbf{K}_\Phi^{-1}\boldsymbol{\Phi}^{\mathrm{T}}\Delta\mathbf{K})^{m-1}\boldsymbol{u}_0 \end{bmatrix} \\ &= \begin{bmatrix} \boldsymbol{r}_1 & \mathbf{C}\boldsymbol{r}_1 & \mathbf{C}^2\boldsymbol{r}_1 & \ldots & \mathbf{C}^{m-1}\boldsymbol{r}_1 \end{bmatrix}.\end{aligned} \quad (20)$$

However, the CARM employed directly might produce singularity, resulting in a large solution error. Both the Gram–Schmidt procedure[6] and the principal component analysis (PCA) method[34] are good choices to make the basis vectors orthogonal to each other to eliminate singularities, and PCA also shows the advantage of enrichment characteristics[35]. The PCA method in this study uses singular value decomposition (SVD) to reconstruct the CARM,

$$\mathbf{R} = \tilde{\mathbf{R}}\boldsymbol{\Sigma}\mathbf{V}^{\mathrm{T}}. \quad (21)$$

The desired CARM is replaced by $\tilde{\mathbf{R}}$ in Equation (21). Summarizing the above formulas, the equilibrium equation can be modified as:

$$\mathbf{K}\tilde{\mathbf{R}}\boldsymbol{y} = \boldsymbol{f}. \quad (22)$$

Multiplying both sides by $\tilde{\mathbf{R}}^{\mathrm{T}}$ to the left, Equation (22) becomes

$$\tilde{\mathbf{R}}^{\mathrm{T}}\mathbf{K}\tilde{\mathbf{R}}\boldsymbol{y} = \tilde{\mathbf{R}}^{\mathrm{T}}\boldsymbol{f}. \quad (23)$$

To simplify the representation, Equation (23) can be written as:

$$\mathbf{K}_{\tilde{\mathbf{R}}}\boldsymbol{y} = \boldsymbol{f}_{\tilde{\mathbf{R}}}, \quad (24)$$

with

$$\tilde{\boldsymbol{u}} = \tilde{\mathbf{R}}\boldsymbol{y}, \quad (25)$$

where $\mathbf{K}_{\tilde{\mathbf{R}}} \in \mathbb{R}^{m \times m}$ and $\boldsymbol{f}_{\tilde{\mathbf{R}}} \in \mathbb{R}^{m}$. $m$ is specified by the users, usually not more than ten, far less than the dimension of the equilibrium equation, which is usually on the order of millions.

# 3. Topology optimization using adaptive auxiliary reduced model reanalysis

The AARMR method reduces the equilibrium equation from an *n*-dimensional linear system to *m*-dimensions, and the computational burden caused by matrix decomposition is effectively alleviated, thereby significantly improving the computational efficiency. This is the motivation for proposing the AARMR method and integrating it into the topology optimization framework.

## 3.1 The process description of the proposed method

Equation (3) should be modified as:

$$\begin{aligned}
\min \quad & : C = \boldsymbol{u}^T \mathbf{K}(\tilde{\boldsymbol{\rho}})\boldsymbol{u}, \\
s.t. \quad & : \sum v_e \tilde{\rho}_e - V \leq 0, \\
& 0 \leq \rho_e \leq 1, \quad e = 1,2,...,N, \\
within \quad & : \tilde{\mathbf{R}}^T \mathbf{K}(\tilde{\boldsymbol{\rho}})\tilde{\mathbf{R}}\boldsymbol{y} = \tilde{\mathbf{R}}^T \boldsymbol{f}, \\
& \boldsymbol{r}_1 = \tilde{\boldsymbol{u}}_0, \\
& \boldsymbol{r}_i = -\tilde{\mathbf{K}}_0^{-1}(\tilde{\boldsymbol{\rho}}_0)\Delta\mathbf{K}(\tilde{\boldsymbol{\rho}},\tilde{\boldsymbol{\rho}}_0)\boldsymbol{r}_{i-1}, \quad i = 2,3,...,m,
\end{aligned} \quad (26)$$

where $\tilde{\mathbf{K}}_0^{-1}(\tilde{\boldsymbol{\rho}}_0)$ is the approximate inverse of the reference stiffness matrix. The initial displacement vector $\tilde{\boldsymbol{u}}_0$ directly utilizes the last solution instead of solving the initial equilibrium equation.

---

**Algorithm 1**: CARM-PARM process

| | |
|---|---|
| 1: | *Determine whether the current iteration index is greater than the activation parameter* <br> ***if*** *loop* < $N_{on}$: |
| 2: | $\boldsymbol{u} \leftarrow MGCG(\mathbf{K}, \boldsymbol{f})$, *Save the displacement solution when loop is close to $N_{on}$, and complete the construction of the PARM when loop is equal to $N_{on}$* |
| 3: | ***else***: |
| 4: | *Construct CARM*: $\mathbf{R} \leftarrow AARMR(\mathbf{K}, \mathbf{K}_0, \boldsymbol{u}_0, \boldsymbol{\Phi})$ |
| 5: | *Approximate solution*: $(\mathbf{R}^T\mathbf{K}\mathbf{R})\boldsymbol{y} = \mathbf{R}\boldsymbol{f}$ and $\boldsymbol{u} = \mathbf{R}\boldsymbol{y}$ |
| 6: | *Determines whether the approximate solution satisfies the criterion* <br> ***if*** $\|\mathbf{K}\mathbf{R}\boldsymbol{y} - \boldsymbol{f}\| / \|\boldsymbol{f}\| < \varepsilon_{tol}$ : |
| 7: | $\boldsymbol{u}$ *is feasible* |
| 8: | ***else*** : |
| 9: | $\boldsymbol{u}_{new} \leftarrow MGCG(\mathbf{K}, \boldsymbol{f})$ |
| 10: | $\boldsymbol{\Phi}_{new} \leftarrow PARM(\boldsymbol{\Phi}_{old}, \boldsymbol{u}_{new})$ |

The flowchart of the entire nested optimization process is shown in Figure 1. In order to make it easier to understand when and how CARM and PARM are performed and updated, the pseudo code of the calculation process in the dotted line wireframe is given in Algorithm 1.

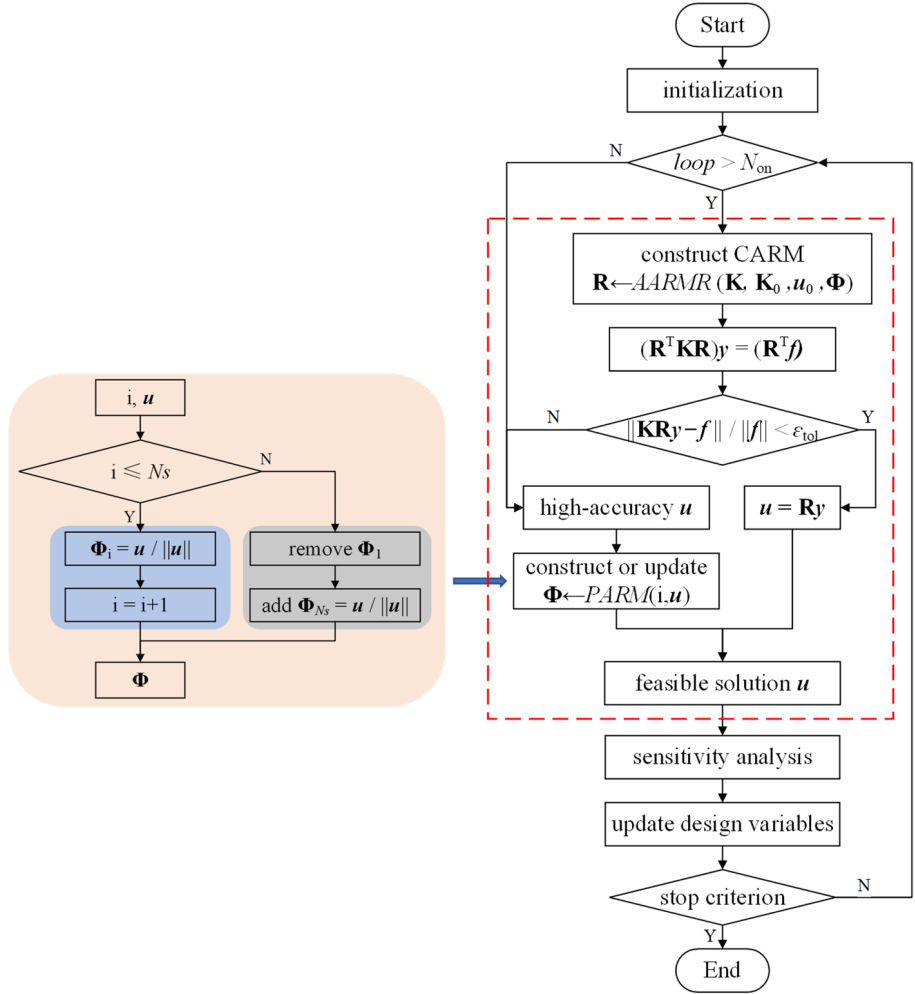

Figure 1 Flowchart of adaptive auxiliary reduced model reanalysis based-topology optimization, the left side of the figure represents the construction or update of the PARM, in which the blue block represents the construction, the gray block represents the update, and $i$ is used to determine the column where the basis vector is located, starting with 1.

It is not suggested to perform the AARMR method at the early optimization stage when the structural topology has undergone huge changes. Therefore, the authors introduce an activation parameter $N_{on}$ to control the activation of CARM, the CARM method only works if the design cycles are more than $N_{on}$. It should be emphasized that the displacement solutions need to be saved when the design cycles close to $N_{on}$ and the construction of the PARM should be completed before using the CARM. Inspired by [6], the construction of the PARM in Equation (20) is based on the previous displacement solutions,

$$\mathbf{\Phi} = [\varphi_1(\boldsymbol{u}_1), \varphi_2(\boldsymbol{u}_2), ..., \varphi_s(\boldsymbol{u}_s)]. \tag{27}$$

According to the experiment, there is no singularity in applying normalized PARM even without orthogonalization. Therefore, the normalized displacement vectors are directly applied to save computational resources,

$$\varphi_i(\boldsymbol{u}_i) = \frac{\boldsymbol{u}_i}{\|\boldsymbol{u}_i\|}. \tag{28}$$

The displacements of the current stage are obtained by MGCG, and PARM is under construction but not applied by CARM, so it is indicated by the blue line. Also marked blue indicates that the updated PARM will be used in the next design cycle.

The CARM is enabled after that the design cycle is greater than $N_{on}$. If the force residual norm (by Equation (29)) does satisfy a given criterion ($\varepsilon < \varepsilon_{tol}$), the approximate solution is considered feasible. In contrast, it is considered that the low-dimensional space based on CARM no longer effectively represents the original solution space, that is, the CARM has been distorted and the solution is infeasible,

$$\varepsilon = \frac{\|\mathbf{K}\tilde{\mathbf{R}}\boldsymbol{y} - \boldsymbol{f}\|}{\|\boldsymbol{f}\|}. \tag{29}$$

---

**Algorithm 2**: two-grid multigrid method with V-cycle.
**Input**: global stiffness matrix $\mathbf{K}$, load vector $\boldsymbol{f}$, initial displacement vector $\boldsymbol{u}_0$, restriction matrix $\mathbf{P}^T$, coarsening level of grid $k$;

---

1:    $v_1$ pre-smooth:    $\boldsymbol{u} = \boldsymbol{u} + \omega \mathbf{D}^{-1}(\boldsymbol{f} - \mathbf{K}\boldsymbol{u}_0)$
2:    coarse grid:    $\boldsymbol{e} = \mathbf{P}^T(\boldsymbol{f} - \mathbf{K}\boldsymbol{u})$
3:    coarse grid solve:    $(\mathbf{P}^T\mathbf{K}\mathbf{P})\boldsymbol{u}_c = \boldsymbol{e}$
4:    interpolate:    $\boldsymbol{u} = \boldsymbol{u} + \mathbf{P}\boldsymbol{u}_c$
5:    $v_2$ post-smooth:    $\boldsymbol{u} = \boldsymbol{u} + \omega \mathbf{D}^{-1}(\boldsymbol{f} - \mathbf{K}\boldsymbol{u})$
6:    Return $\boldsymbol{u}$

---

This is because the PARM does not match the current structural topology, resulting in an inaccurate approximate inverse. This is the essence of the "adaptive" concept, updating PARM to improve the accuracy of the approximate inverse and thus prolong the fidelity period of CARM. Updating PARM is similar to *on-the-fly* ROM[6], the earliest basis vector should be removed and the latest normalized displacement vector should be added. However, both the construction and updating of PARMs require highly accurate solutions, the solution of equilibrium equations by direct solvers is still a considerable burden. Sequentially, the multi-grid conjugate gradient (MGCG) method[5] is used to obtain highly accurate solutions to improve the efficiency of the PARM. The essence of MGCG is to employ the multigrid method as the preconditioner of the PCG method, which is described in detail in Algorithm 2.

The CARM is allowed to be constructed dynamically based on the PARM to reduce the construction cost. In each iteration, the last stiffness matrix is used to reconstruct the CARM to reduce the modification amount as much as possible. In addition, the selection of the initial displacement is also related to the solution efficiency and accuracy of CARM. The approximate solution of the previous design cycle ($u_{last}$) is directly used as the initial solution in the current iteration, instead of multiplying the load vector by the approximate inverse of the reference stiffness matrix ($\mathbf{\Phi} \mathbf{K}_\Phi^{-1} \mathbf{\Phi}^T f$). The study shows that this method reduces a large number of matrix multiplications and has little effect on the results. The results based on the example in Section 4.1.1 are shown in Figure 2. Both the relative differences in the force residual norm and compliance are sufficiently small to be neglected.

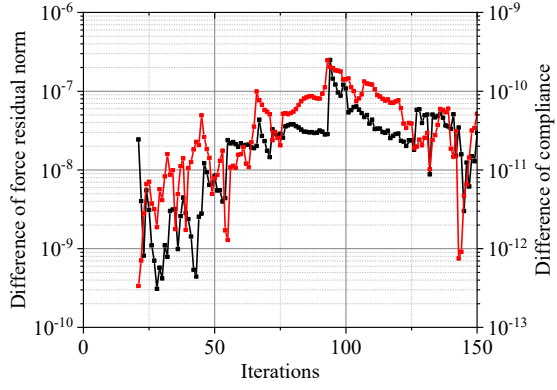

Figure 2 Relative differences of force residual norm (red line) and compliance (black line) with different initial displacement choices.

It should be noted that the curve is not calculated from the beginning of the iteration because of the introduction of the activation parameter. The activation parameter $N_{on}$ is usually no more than a dozen but must be greater than the number of basis vectors of PARM.

However, in some topology optimization cases, sudden changes can also occur in the middle stages of the optimization, an extreme case is designed to find out the resistance of AARMR to this situation. The volume fraction changes from 0.48 at the 50th iteration and decreases by 0.005 per iteration, until it decreases to 0.45 (at the 56th iteration) and then remains constant. The optimization process is shown in Figure 3, an indicator is created to evaluate the changes in the density field,

$$w = \frac{\sum_{i=1}^{N}\left|x_i^{(k+1)} - x_i^{(k)}\right|}{N} \times 100\%, \tag{30}$$

where $k$ denotes the iteration index and $N$ denotes the total number of elements. The change in volume fraction causes a large change in the density field, yet the increase of $w$ is not large enough to destabilize the optimization process, and this perturbation can also lead to the creation of new members of the structure. Furthermore, as shown by the variation of the force residual curve, the fidelity of AARMR returns to normal in just a dozen iterations. Therefore, AARMR can also maintain its efficiency and may perform better for general topology optimization with more moderate changes in the middle stages.

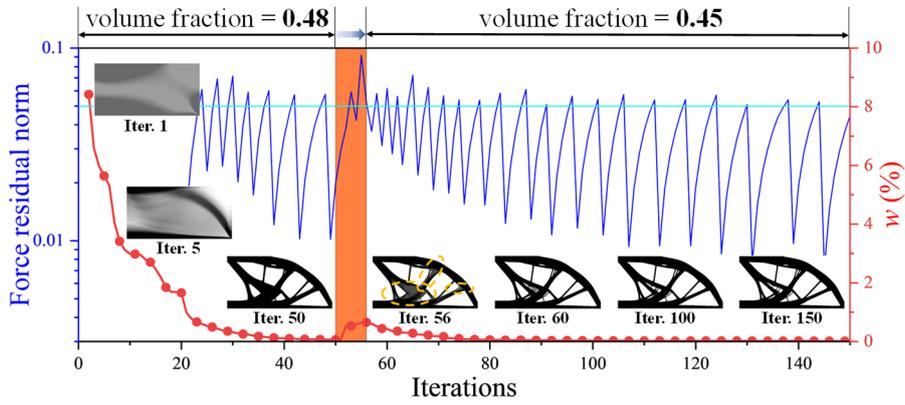

(a) The geometric model (640×320 grid) and boundary conditions.

(b) Curves of force residuals and changes of the iterative process.

Figure 3 Iterative process considering the change of volume fraction. The activation parameter $N_{on}$ is 20, $Ns = 2$, $Nm = 2$, the maximum CG iterations for MGCG are 200, the convergence tolerance is set to $10^{-6}$, and the grid level is 3.

## 3.2 Filter strategy and sensitivity analysis

Density filter technology[29] is selected to suppress the checkerboard phenomenon.

$$\tilde{\rho}_i = \frac{\sum_{j \in N_i} \omega(\boldsymbol{x}_j) \rho_j}{\sum_{j \in N_i} \omega(\boldsymbol{x}_j)}, \tag{31}$$

where the weighting function $\omega(\boldsymbol{x}_j)$ is a linearly decaying (cone-shape) function with the distance between elements as a variable,

$$\omega(\boldsymbol{x}_j) = \max(0, r - \|\boldsymbol{x}_i - \boldsymbol{x}_j\|). \tag{32}$$

Unless otherwise specified, the filter radius $r$ is 2.5 times the element length, both for 2D and 3D scenarios. When the filter strategy is determined, the sensitivity can be determined by the chain rule,

$$\frac{\partial C(\tilde{\boldsymbol{\rho}})}{\partial \rho_e} = \sum_{i \in N_e} \frac{\partial C(\tilde{\boldsymbol{\rho}})}{\partial \tilde{\rho}_i} \frac{\partial \tilde{\rho}_i}{\partial \rho_e} = \sum_{i \in N_e} \frac{\omega(\boldsymbol{x}_i)}{\sum_{j \in N_i} \omega(\boldsymbol{x}_j)} \frac{\partial C(\tilde{\boldsymbol{\rho}})}{\partial \tilde{\rho}_i}, \qquad (33)$$

where the $\partial C(\tilde{\boldsymbol{\rho}})/\partial \tilde{\rho}_i$ employ approximate sensitivity and is given as:

$$\frac{\partial C(\tilde{\boldsymbol{\rho}})}{\partial \tilde{\rho}_i} = -\boldsymbol{y}^{\mathrm{T}} \tilde{\mathbf{R}}^{\mathrm{T}} \frac{\partial \mathbf{K}}{\partial \tilde{\rho}_i} \tilde{\mathbf{R}} \boldsymbol{y}. \qquad (34)$$

## 4. Numerical examples

In this section, several classic structural topology optimization examples are employed to verify the method proposed in this study. Using MGCG[5] as a benchmark method, the computational efficiency improvement in minimizing end-compliance is compared. To compare the efficiency, the speedup can be defined as:

$$speedup = \frac{T_{\mathrm{Ref}}}{T_{\mathrm{Goal}}}, \qquad (35)$$

where $T_{\mathrm{Ref}}$, $T_{\mathrm{Goal}}$ denote the cumulative time for solving the equilibrium equation employing MGCG and AARMR, respectively. The elastic parameters: Young's moduli $E = 1$ and $E_{\min} = 10^{-9}$, Poisson's ratio $v = 0.3$, and penalty factor $p = 3$ were applied in both cases. All experiments run on a PC with an Intel Core i7-8750H @2.6 GHz with 32 GB of RAM.

### 4.1 Influences of model parameters on the efficiency

The operations required to construct reduced models are related to the terms of basis vectors, and the grid level of MGCG also contributes to the computational cost. Therefore, the influence of model parameters on the efficiency is discussed, which provides guidance for the subsequent comparison of computational efficiency of methods. The MGCG outputs approximate displacements when the force residual norm does not meet the tolerance or the maximum CG iterations is reached. It should be emphasized that although both AARMR and MGCG use the force residual norm as a certain standard, their application scenarios are different. The former is the criterion for updating the PARM, while the latter is the criterion for controlling the accuracy of the output results. To make it easier to distinguish between the two, the convergence tolerance specifically refers to the force residual norm used at MGCG.

### 4.1.1 Terms of basis vectors

The optimization object is a 2D cantilever beam sketched in Figure 4. The design domain is discretized into a 300×180 finite element (FE) grid with linear quadrilateral plane-stress elements of size 1×1, the whole left side is fully restrained, and a static force is loaded at the midpoint of the right end. It is assumed that the available material can only cover 50% of the volume of the design domain and $\varepsilon_{tol}$ is 0.1. The optimization process involves MGCG, whose convergence conditions are set with reference to the literature [5], the convergence tolerance is set to $10^{-6}$, the maximum CG iterations is 200, and the grid level is 3.

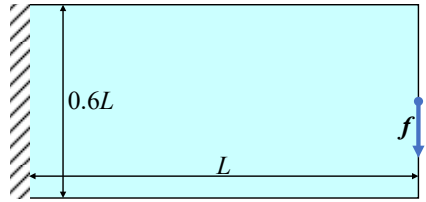

Figure 4 Geometrical settings to minimize the end-compliance of the 2D cantilever beam.

In addition to the cost of constructing reduced models, the accuracy of optimization results might also be related to the number of linearly independent basis vectors. Therefore, the relationship between the number of basis vectors (both PARM (*Ns*) and CARM (*Nm*)) and efficiency improvement and accuracy of optimization results should be determined. *Ns* and *Nm* are set to [1, 2, 3, 4, 6, 8, 10] respectively, for a total of 49 test cases. As shown in Figure 5, the results illustrate that the minimum cumulative computational cost is 20.90 with *Ns* of 2 and *Nm* of 2, but the maximum is as high as 28.34 with *Ns* of 1 and *Nm* of 10.

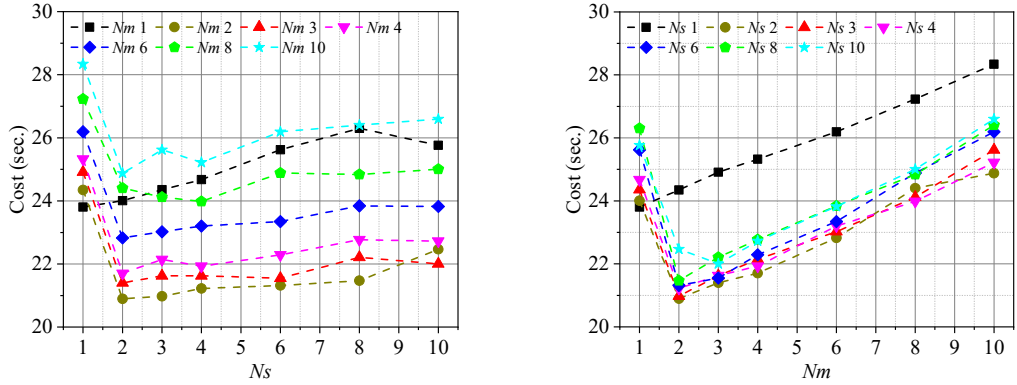

(a) Cumulative computational cost for *Ns*  (b) Cumulative computational cost for *Nm*

Figure 5 Influences of *Ns* and *Nm* on cumulative computational cost.

The curves of cumulative computational cost have similar patterns of variation with respect to *Ns* or *Nm*. Cumulative computational costs decrease first (from 1 to 2) and then increase (from 2 to 10), and it has a significant positive linear correlation with the terms of basis vectors in the increasing process. However, the changes of the lines in Figure 5(a) are slower and intervals between the lines are larger, whereas the changes of the lines in Figure 5(b) are more obvious and the lines are relatively concentrated. Therefore, the terms of basis vectors have a great influence on computational efficiency, especially for *Nm*. It can be seen from the test results that the minimum value always occurs when *Ns* is in the range of [2, 4] and *Nm* = 2.

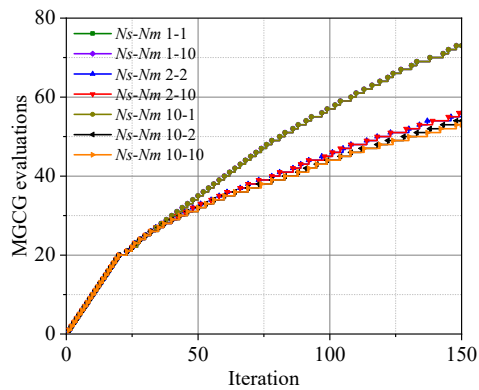

Figure 6 Influences of Ns and Nm on cumulative times of MGCG evaluations.

Lines *Ns-Nm* "1-1", "1-10" and "10-1" in Figure 5 are completely overlap, and there are slight differences between the remaining lines. Figure 6 illustrates that the selection of *Ns* and *Nm* slightly affects the MGCG evaluations unless one of them is 1. Note that there is almost consistency between lines *Ns-Nm* "2-2" and "2-10" and between lines *Ns-Nm* "10-2" and "10-10". It indicates that *Ns* might be a more important effect on the evaluations of MGCG, although the difference is small.

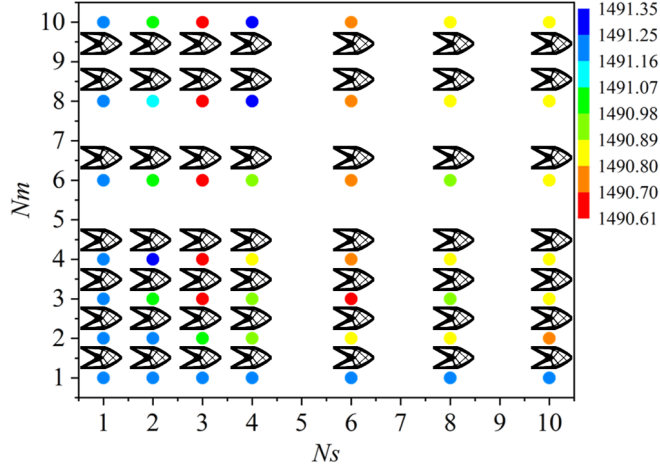

Figure 7 Optimization results with different combinations of *Ns* and *Nm*.

Above discussion reveals that the optimization process is not sensitive to the selection of *Nm* and *Ns*, and the difference in computational cost should be attributed to the construction of the reduced model. The CARM is reconstructed at each iteration, but the PARM only needs to be updated when the force residual criterion is violated. In other words, the frequency of reconstructing CARM is higher during the entire optimization process. Therefore, it is recommended to take a small *Nm*. It should be noted that feasible solutions can be obtained when *Ns* or *Nm* is 1, which reflects the excellent performance in capturing boundary conditions. The optimization results of all cases are shown in Figure 7, the structural characteristics of the optimization results tend to be the same. The discrepancy between different parameter settings is also reflected in terms of end-compliance, it is distinguished by different colors according to the level of end-compliance. Although the end-compliance varies from case to case, the magnitude of variation is extremely small, the relative difference of end-compliance is calculated by Equation (36), their maximum difference of end-compliance is only

0.049%.

$$diff = \frac{C_{\text{Goal}} - C_{\text{Ref}}}{C_{\text{Ref}}} \times 100\%, \quad (36)$$

where $C_{\text{Goal}}$, $C_{\text{Ref}}$ denote the end-compliance employing AARMR and MGCG, respectively. To display the optimization results more clearly, the structures of the enlarged display size is given in Figure 8, the *Ns* vary from left to right in ascending order and from top to bottom *Nm* in descending order.

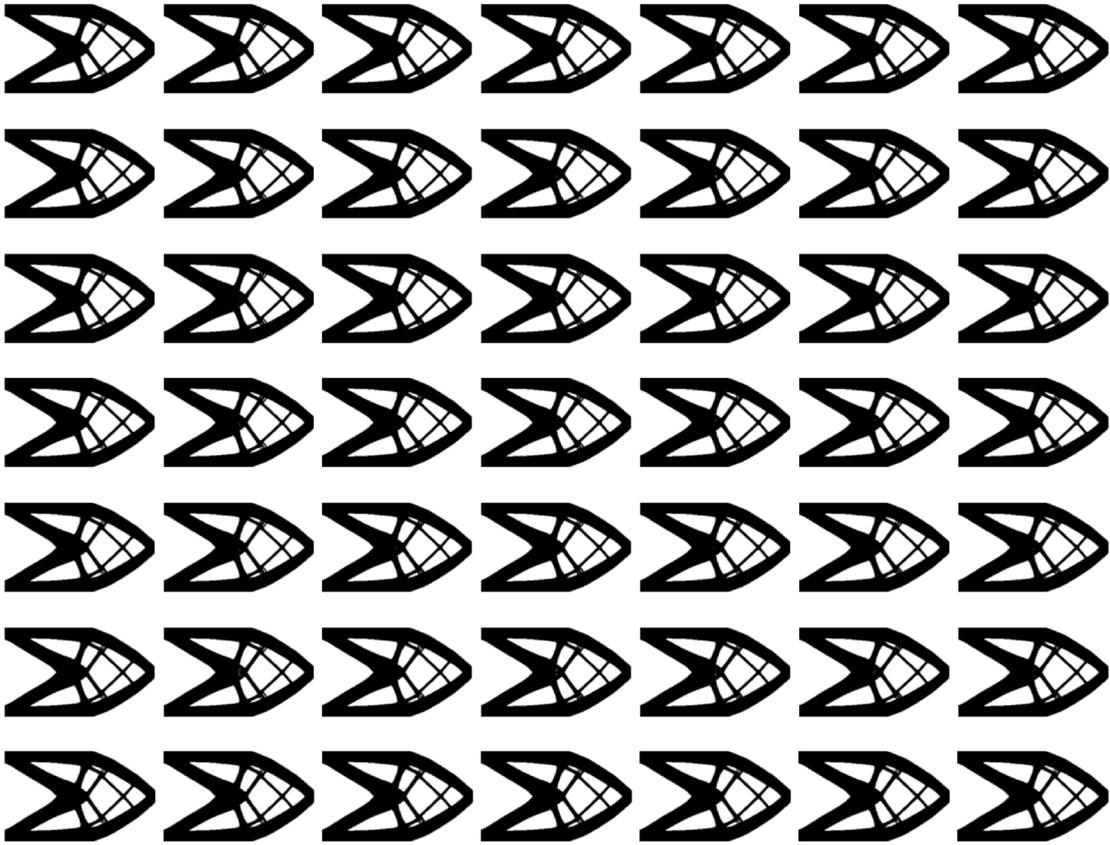

Figure 8 Structural topologies with different combinations of *Ns* and *Nm*.

The end-compliance with *Ns* of 3 and *Nm* of 8 is the smallest, the end-compliance with *Ns* of 4 and *Nm* of 10 is the largest. Comparison of the differences between the two density fields reveals only some local differences (as shown in Figure 9). These differences occur in a very small area of the entire structure, and most of them are relatively small. Therefore, it can be inferred that the solution may be insensitive to *Ns* or *Nm*.

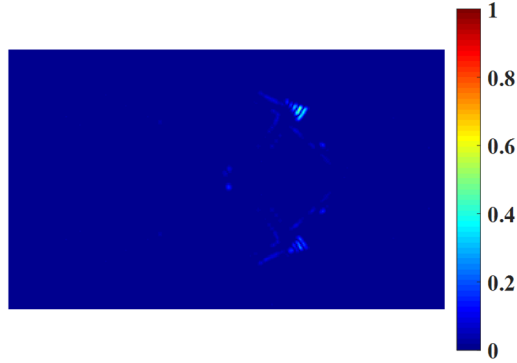

Figure 9 Absolute difference of two density fields

**4.1.2 Grid level of V-cycle**

As mentioned in [5], the grid level has an influence on computational efficiency. Increasing the coarsening level means that the dimensionality of the linear system at the coarsest grid is reduced, but the more matrix-vector product operations are required. The design domain and boundary conditions of the optimization objective are consistent with Sec 4.1.1. Considering that the grid levels are 2, 3 and 4, *Ns* and *Nm* are both set to 2, and the rest of the parameter settings are also consistent with Sec 4.1.1.

The number of CG iterations that required for each MGCG are counted, in which the AARMR method calls MGCG intermittently, so it is represented by dots. As shown in Figure 10, whether calling MGCG in AARMR, or using MGCG continuously (Subsequently, AARMR and MGCG represent these two modes respectively), the more grid levels, the more CG iterations are required. However, there is no direct relationship between the two solving modes and the average CG iterations. When the grid level is 2 or 3, the average CG iteration of AARMR (9.07 for level 2 and 9.61 for level 3) is greater than that of MGCG (7.63 for level 2 and 9.04 for level 3), otherwise it is vice versa (11.30 for AARMR with level 4 and 13.61 for MGCG with level 4).

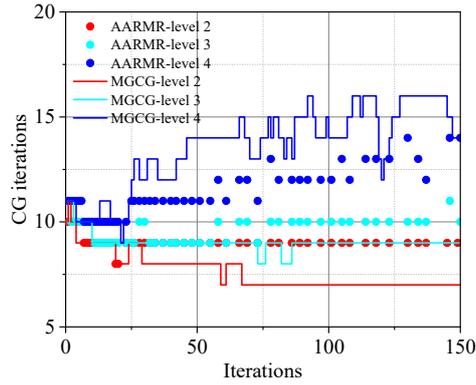

Figure 10 The influence of grid level on CG iterations. AARMR-level indicates the count value when AARMR calls MGCG, and MGCG-level indicates that MGCG is always used.

The difference in CG iterations is also reflected in the computational cost. The 88-line code[31] based on direct solver is supplemented as a reference, and the acceleration effect is shown in Figure 11. There is no linear correlation between grid level and computational cost, which is a trade-off between the complexity of solving linear system and matrix-vector product. When the grid level is 2, the performance is the worst, while the selection of 3-layer or 4-layer grid has almost no influence on computational cost of AARMR, but a greater influence on MGCG. The speedup of AARMR and MGCG relative to 88-line code are the best when the grid layer is 3. However, the speedup of AARMR relative to MGCG is the worst at this time. In addition, although AARMR needs to call MGCG, the frequency of using MGCG is much less than that of the mode of using MGCG continuously, so the selection of grid layers has relatively little influence on computational cost of AARMR.

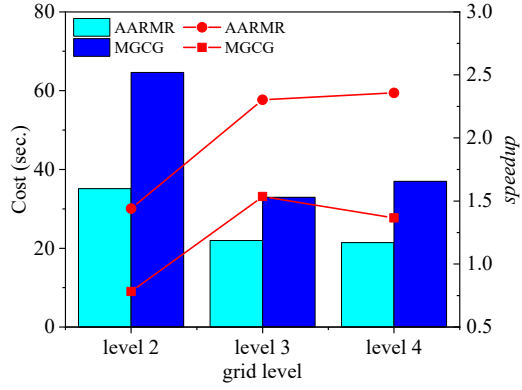

Figure 11 Comparison of computational cost and speedup. The bars represent computational cost, and the line represents speedup.

The topology of the optimization results is given in Figure 12, AARMR is closer to the 88-line code both in terms of graph results and optimization objectives. MGCG outperforms in terms of the stability of optimization results, although the maximum difference in the optimization objectives of AARMR at different grid levels is 0.0087%. In addition, the largest relative difference between AARMR and MGCG is only -0.34%.

The test results show that the grid level significantly affects the computational cost and has little effect on the accuracy of the optimization results. Comparing AARMR with MGCG is more representative when the grid level is 3, because the speedups of both relative to the 88-line code are excellent in this scenario, and the speedups of AARMR and MGCG are closest. Therefore, it makes more sense to set the grid level to 3 in the subsequent comparison of speedup.

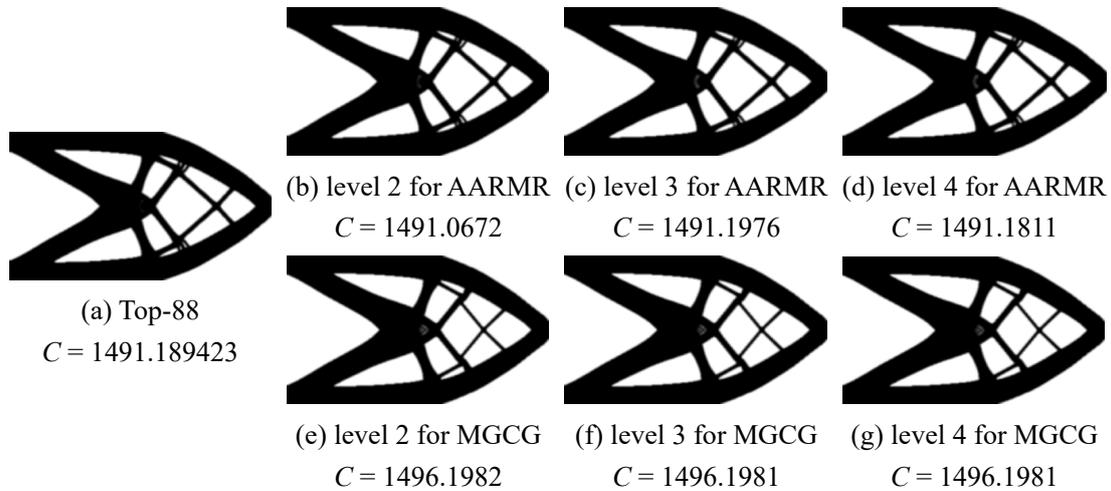

(a) Top-88
$C = 1491.189423$

(b) level 2 for AARMR
$C = 1491.0672$

(c) level 3 for AARMR
$C = 1491.1976$

(d) level 4 for AARMR
$C = 1491.1811$

(e) level 2 for MGCG
$C = 1496.1982$

(f) level 3 for MGCG
$C = 1496.1981$

(g) level 4 for MGCG
$C = 1496.1981$

Figure 12 The optimized structure topology with different grid levels.

**4.1.3 Memory requirement**

To better reflect the requirements of the proposed method on the device memory, the memory usage records of the iterative process are intercepted for analysis. MGCG is a well-performing method that requires less memory than the direct method, especially for 3D problems. Therefore, it is reasonable to employ MGCG as a reference to evaluate the memory dependence of the proposed method. Without loss of generality, the direct solver is also employed for comparison and tested for both 2D and 3D problems (The design domain and boundary conditions are from Sec4.1.1 and 4.2.2, respectively). The grid level is set to 3, $Ns$ and $Nm$ are both set to 2. The segment of the memory usage records of their iterative process is shown in Figure 13.

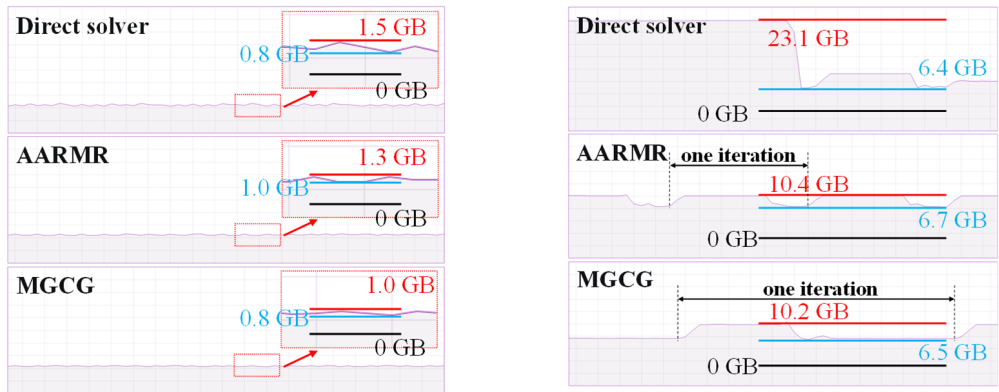

(a) 2D FE model with 640×384 grid    (b) 3D FE model with 108×36×72 grid

Figure 13 Comparison of memory requirements in 2D and 3D problems.

The memory usage when the program is not running is used as the starting evaluation point (0GB) due to the presence of other software. Compared to 3D problem, 2D problem require much less memory, resulting in a less obvious period of fluctuation. In a complete iteration, there is a peak and a trough, where the peak is caused by solving the equilibrium equations and the trough is the memory occupation of the variables during the iteration. The troughs of the direct solver and MGCG are similar and smaller than that of AARMR, which is due to the fact that AARMR employed MGCG intermittently and generates more variables. In general, the direct solver has the highest memory usage, which is more significant in the 3D problem. The memory usage increase caused by more variables did not further widen the gap between AARMR and MGCG. In addition, the memory fluctuations of the 3D problem reflect an iterative period, and it is clear that the period of AARMR is shorter and thus more efficient. The detailed efficiency comparison will be discussed next.

## 4.2 Scaling of speedup in different scenarios

In this section, the acceleration effect of minimizing end-compliance in different scenarios is considered, while observing the influence of the given force residual criterion on the optimization results and iterative processes. The optimization objects are replaced with a 2D half-wheel sketched in Figure 14 and a 3D simply supported beam sketched in Figure 21. In the following, MGCG is used as the benchmark method.

### 4.2.1 2D scenario: half-wheel

The design domain is a rectangle with an aspect ratio of 2 and is discretized into FE models of different scales with linear quadrilateral plane-stress elements of size 1×1. Four FE models with increasing order of resolution are considered: Model 1(320×160), Model 2(640×320), Model 3(840×420), Model 4(960×480). The boundary conditions of all FE models are consistent, whose lower-left corner is fully constrained, and the DOFs of the load direction in the lower-right corner are constrained. A static force is loaded at the midpoint of the bottom. The volume fraction is set to 0.5, activation parameter $N_{on}$ is 20, $Ns = 2$, and $Nm = 2$. The maximum CG iterations for MGCG are 200, the convergence tolerance is set to $10^{-6}$, and the grid level is 3.

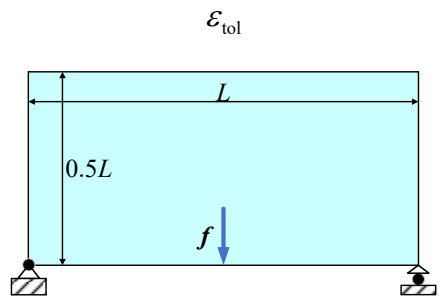

$\varepsilon_{tol}$

Figure 14 Geometrical settings to minimize the end-compliance of the 2D half-wheel.

The speedup with different force residual criteria $\varepsilon_{tol}$ is shown in Figure 15, the maximum speedup can reach 2.32 when $\varepsilon_{tol}$ is equal to 5%. In general, the speedup tends to increase as the problem scale increases. This indicates the potential of the AARMR method in saving computational costs, especially for large-scale problems. It can be seen intuitively that the acceleration effect becomes worse as the setting of $\varepsilon_{tol}$ becomes stricter, that is, the increase in speedup becomes smaller as the residual criterion is stricter. However, the minimum speedup can also reach 1.26 when $\varepsilon_{tol}$ is equal to 0.5%.

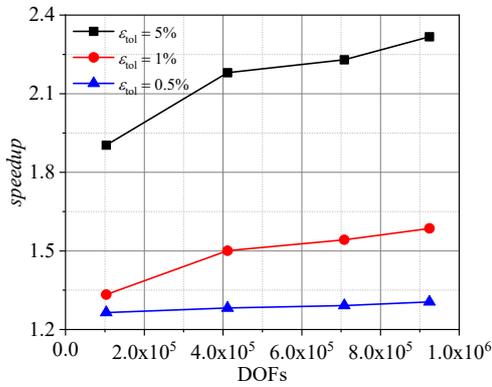 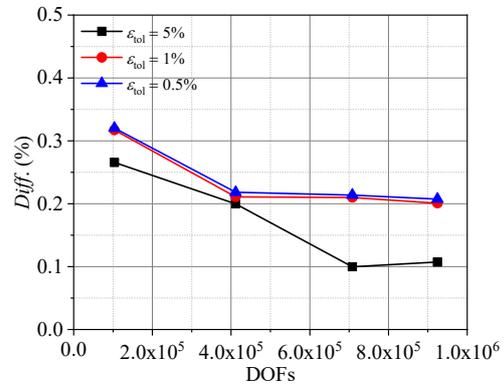

Figure 15 Speedup of the 2D half-wheel case at different scales.    Figure 16 Difference of end-compliance for the 2D half-wheel case.

Figure 18 illustrates that the results for different problem scales with different force residual criteria are almost consistent with the structural topology obtained by MGCG. There are only slight differences in details, and the strict force residual criteria make structural details more consistent. Furthermore, the differences in end-compliance under different $\varepsilon_{tol}$ are controlled within an acceptable range (as shown in Figure 16). The absolute values of the differences between the examples here are of the same order of magnitude as the 2D cantilever test results above, and remain within 0.35%.

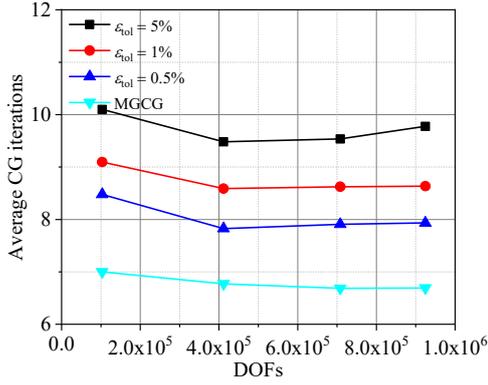
(a) Average CG iterations for MGCG.

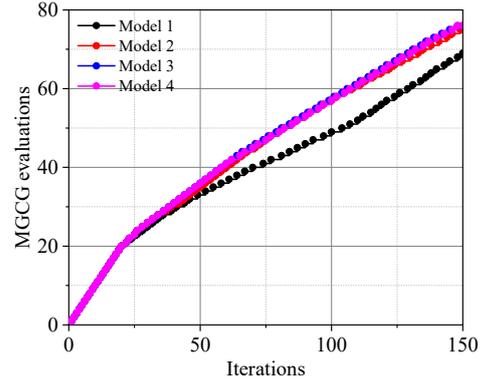
(b) Residual criterion 0.5%.

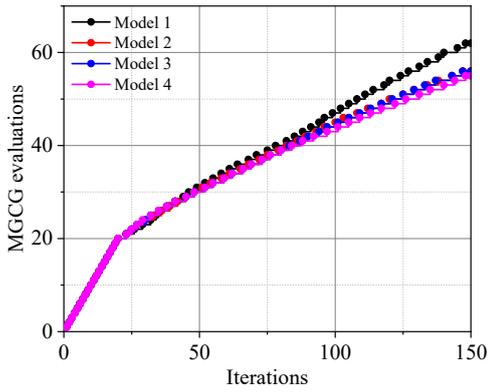
(c) Residual criterion 1%.

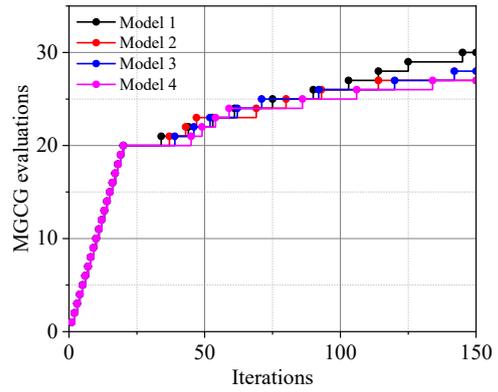
(d) Residual criterion 5%.

Figure 17 Cumulative MGCG evaluations for the 2D half-wheel case.

The reason for the difference in speedup can be clarified according to the cumulative MGCG evaluations, as shown in Figure 17. Decreasing the force residual criterion will result in more MGCG evaluations, while the change of the scale has almost no effect on the MGCG evaluations, especially for large-scale problems. Note that the average CG iterations of AARMR are higher than that of MGCG, so it is inferred that the contribution of speedup of AARMR does not come from calling MGCG, and even part of the computational cost of AARMR calling MGCG is more expensive than that of using MGCG continuously. However, reducing the force residual criterion is conducive to improving the accuracy that the AARMR method can achieve, and the structural topology (given in Figure 18) is closer to the MGCG method.

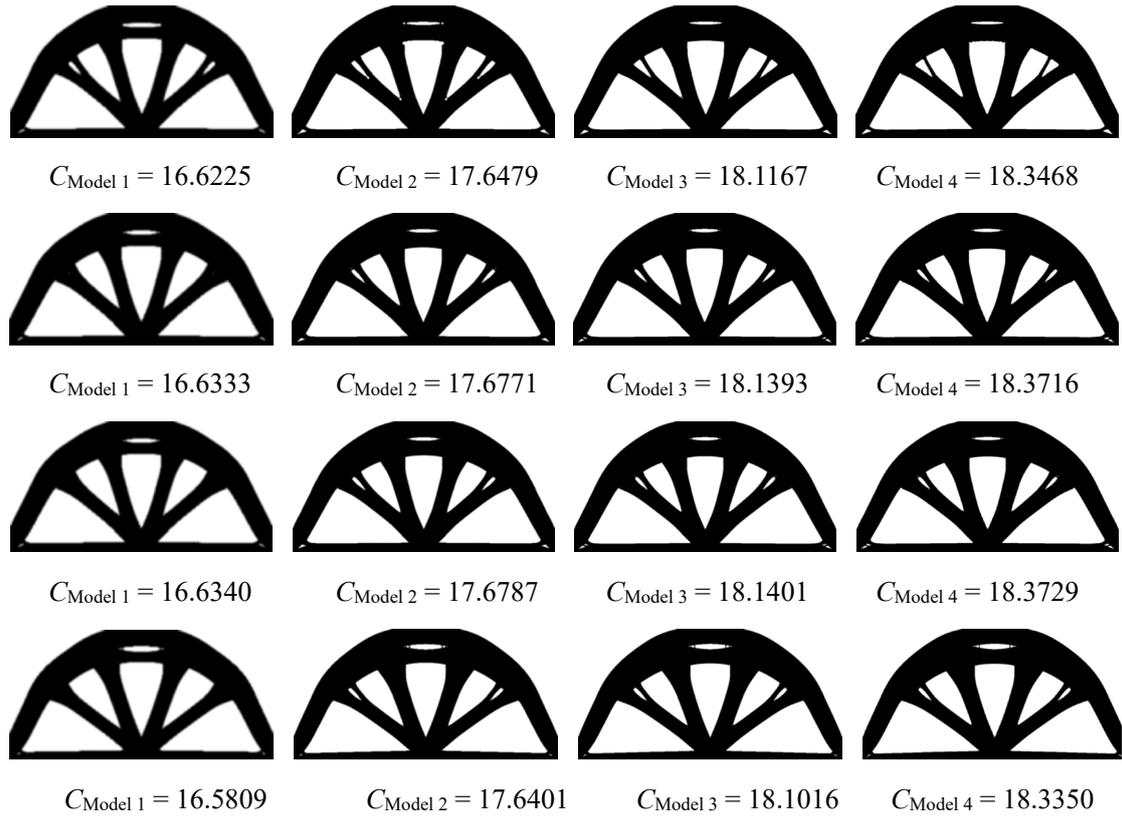

| $C_{\text{Model 1}} = 16.6225$ | $C_{\text{Model 2}} = 17.6479$ | $C_{\text{Model 3}} = 18.1167$ | $C_{\text{Model 4}} = 18.3468$ |
| --- | --- | --- | --- |
| $C_{\text{Model 1}} = 16.6333$ | $C_{\text{Model 2}} = 17.6771$ | $C_{\text{Model 3}} = 18.1393$ | $C_{\text{Model 4}} = 18.3716$ |
| $C_{\text{Model 1}} = 16.6340$ | $C_{\text{Model 2}} = 17.6787$ | $C_{\text{Model 3}} = 18.1401$ | $C_{\text{Model 4}} = 18.3729$ |
| $C_{\text{Model 1}} = 16.5809$ | $C_{\text{Model 2}} = 17.6401$ | $C_{\text{Model 3}} = 18.1016$ | $C_{\text{Model 4}} = 18.3350$ |

Figure 18 Comparison of structural topologies at different scales. The first three rows are AARMR, from top to bottom, with criteria of 5%, 1%, and 0.5%, and the bottom is MGCG.

This paper aims to overcome the application bottleneck of CA in topology optimization, so AARMR and MGCG are not in a competitive relationship. However, to further discuss the difference between AARMR and MGCG, the convergence tolerance of MGCG is set to be consistent with the updated criteria of PARM ($\varepsilon_{\text{tol}}$ is constant at 1%). As shown in Figure 19, the strict convergence tolerance significantly improves the accuracy of the approximate solution of MGCG, while it is slightly less efficient for AARMR. When convergence tolerance is $10^{-6}$, AARMR is able to achieve computational accuracy that is significantly better than the relaxed one.

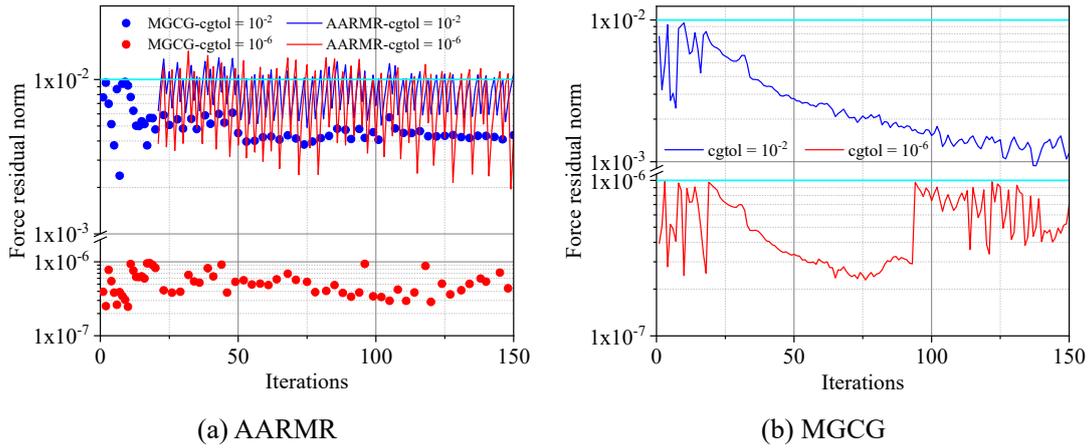

(a) AARMR  (b) MGCG

Figure 19 Force residual iteration curve with different convergence tolerances, cgtol is the convergence tolerance in MGCG, the dots represent the accuracy of the calculation results of calling MGCG when AARMR is not applicable or violates the force residual criteria.

Although the relaxed convergence tolerance results in only one CG iteration for MGCG in the middle and late optimization iterations, AARMR still has a slight advantage over MGCG, computational efficiency increased by 6.4 percentage points. The efficiency of AARMR has been verified from various aspects. However, the purpose of this paper is not to improve MGCG, but to overcome the difficulty of CA in solving large-scale problems. To guarantee the generalization ability of AARMR, the convergence tolerance of MGCG will use a strict one.

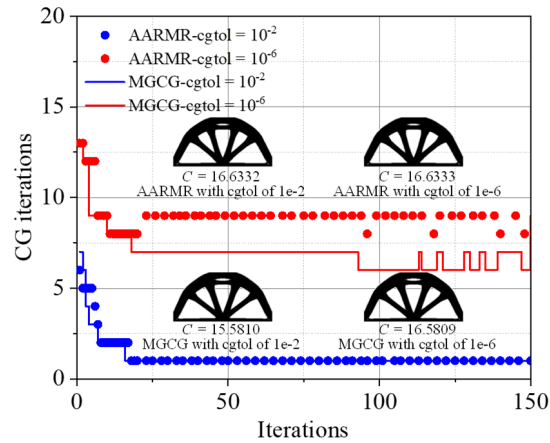

Figure 20 CG iterations with different convergence tolerances.

### 4.2.2 3D scenario: simply supported beam

To further prove the competence of proposed method, 3D problems should be investigated. The design domain is a cuboid and is discretized into FE models of different scales with hexahedral elements of size 1×1×1. Four FE models with increasing order of resolution are considered: Model 1 (grid 72×24×48), Model 2 (grid 84×28×56), Model 3 (grid 96×32×64), and Model 4 (grid 120×40×80). The boundary conditions of all FE models remain consistent, and the four corners at the bottom are completely constrained. A static force is loaded at the midpoint of the bottom. The volume fraction is 0.2, activation parameter $N_{on}$ is 20, $Ns = 2$, and $Nm = 2$. The maximum CG iterations for MGCG are 50, the convergence tolerance is set to $10^{-6}$, and the grid level is 3.

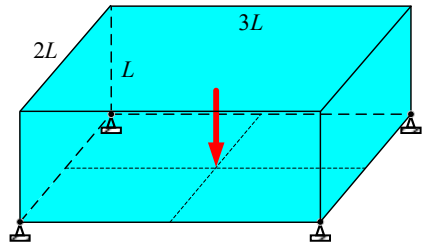

Figure 21 Geometrical settings to minimize the end-compliance of 3D simply supported beam.

Figure 22 illustrates the cumulative computational cost of AARMR versus MGCG at different problem scales. The computational cost of the two methods maintains linear growth in both 2D and 3D scenarios, but the difference in slopes is more prominent in 3D scenarios. The AARMR can alleviate the computational burden for dealing with 3D problems, which is also reflected in the speedup (as shown in Figure 23). As the problem scale increases, the speedup stabilizes within a certain range, almost double that of 2D scenarios. The maximum speedup reached 4.85, and the computational cost might be decreased significantly.

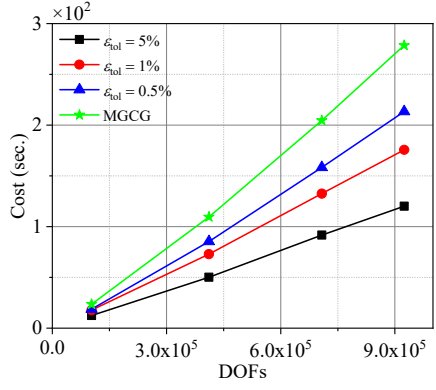 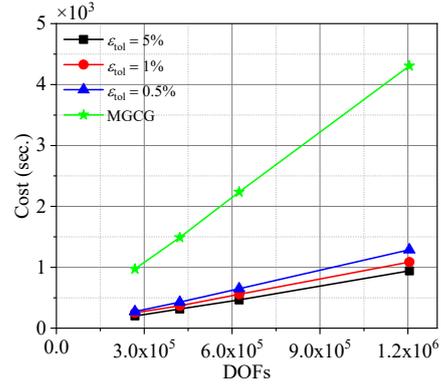

(a) 2D half-wheel.   (b) 3D simply supported beam.

Figure 22 Comparison of cumulative computational cost in different scenarios.

Figure 24 illustrates that the maximum difference is only approximately 0.19%, and strict criteria seem to be more conducive to improving the stability of objectives. It should be stated that the exact solutions are used to compute objectives for 2D problems, while the 3D problems are difficult to be solved accurately due to memory constraints, so the MGCG are used to compute objectives. There is only a slight visually visible difference between MGCG and AARMR in terms of optimized results given in Figure 26. Note that the optimization results are symmetric due to the symmetry of the constraints and loading schemes. To clearly represent the internal structure, Figure 26 also shows a cross-sectional view of MGCG. Whether for 2D problems or 3D problems, the optimized structural topologies of different problem scales are consistent with that of MGCG, indicating that the AARMR method has great computational accuracy.

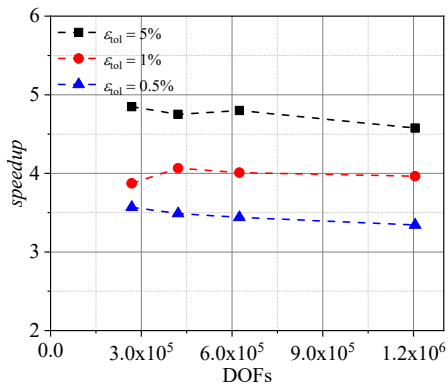 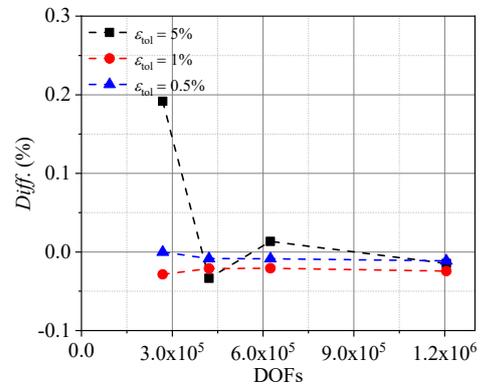

Figure 23 Speedup of the 3D simply supported beam at different scales.   Figure 24 Difference of end-compliance for the 3D simply supported beam.

Similarly, the force residual criterion $\varepsilon_{tol}$ affects speedup because its value is related to the evaluations of the MGCG (as shown in Figure 25). Strict criteria will lead to more MGCG evaluations, and the number of MGCG evaluations may be more sensitive to criteria but less sensitive to the problem scale. Furthermore, 3D scenarios require fewer MGCG evaluations than 2D scenarios under the same force residual criterion. It should be emphasized that the average CG iterations of both AARMR calling MGCG and continuous MGCG reached the limit of 50.

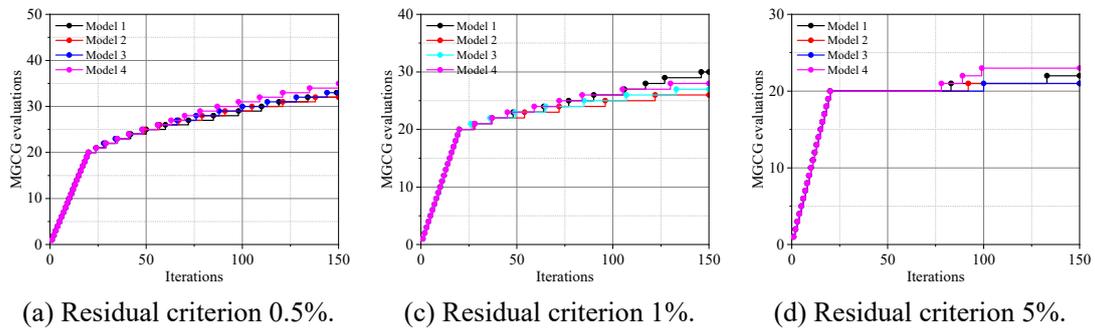

(a) Residual criterion 0.5%.     (c) Residual criterion 1%.     (d) Residual criterion 5%.

Figure 25 Cumulative MGCG evaluations for 3D simply supported beam.

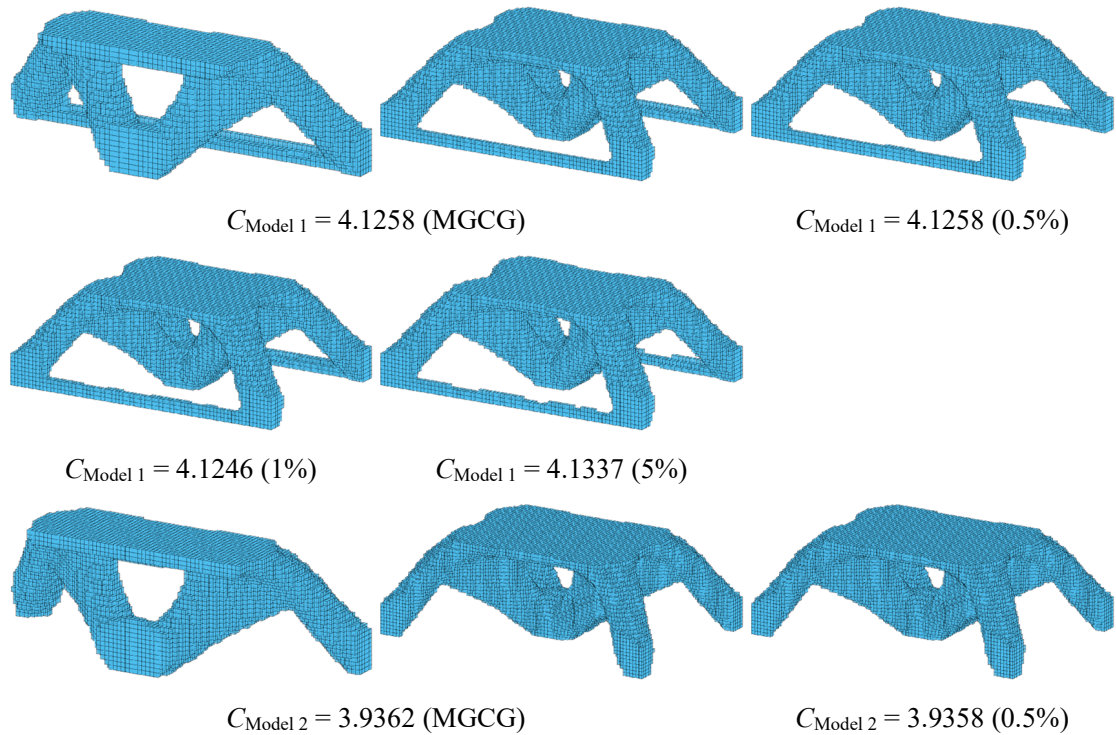

$C_{\text{Model 1}}$ = 4.1258 (MGCG)     $C_{\text{Model 1}}$ = 4.1258 (0.5%)

$C_{\text{Model 1}}$ = 4.1246 (1%)     $C_{\text{Model 1}}$ = 4.1337 (5%)

$C_{\text{Model 2}}$ = 3.9362 (MGCG)     $C_{\text{Model 2}}$ = 3.9358 (0.5%)

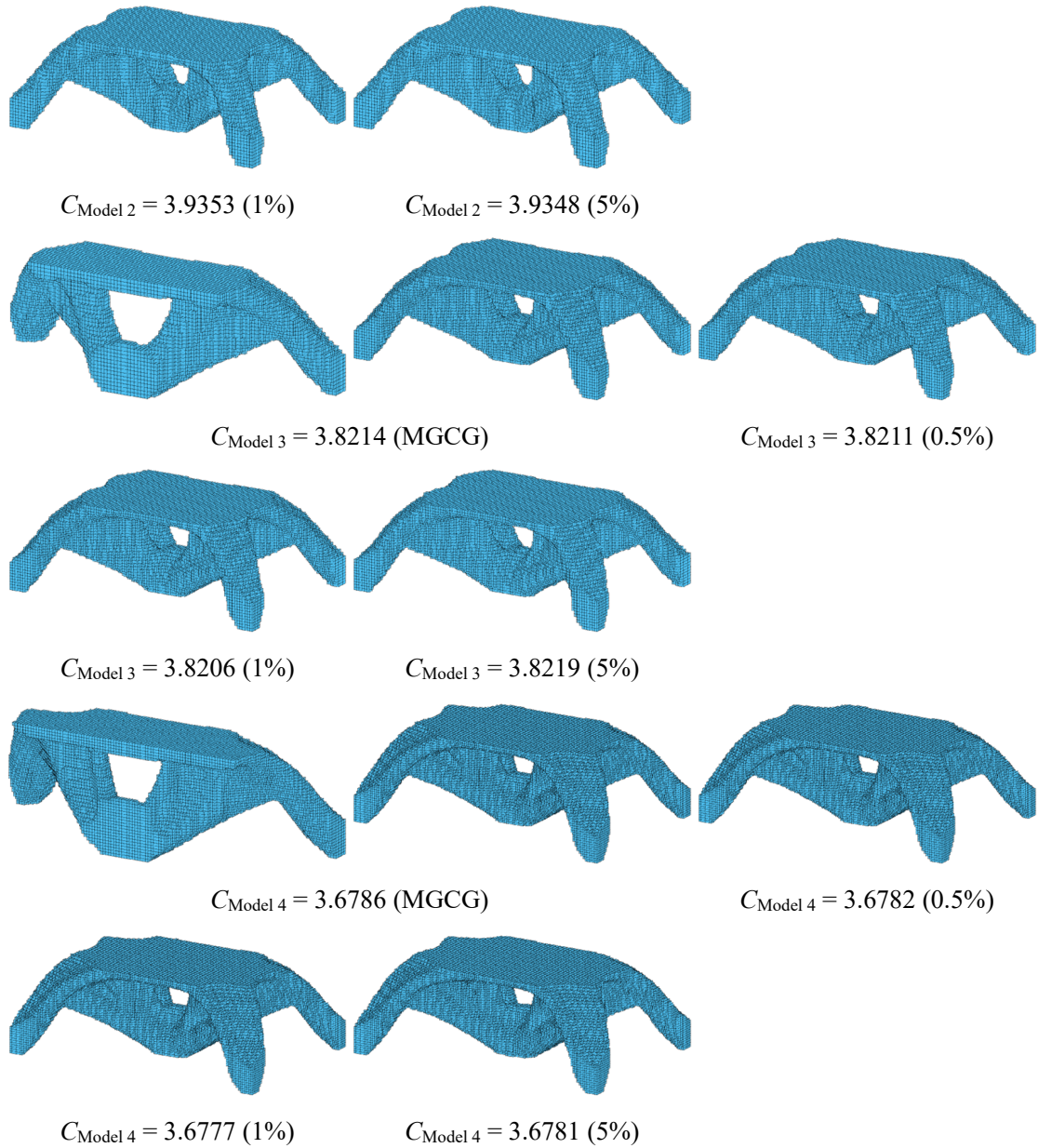

$C_{\text{Model 2}} = 3.9353$ (1%)   $C_{\text{Model 2}} = 3.9348$ (5%)

$C_{\text{Model 3}} = 3.8214$ (MGCG)   $C_{\text{Model 3}} = 3.8211$ (0.5%)

$C_{\text{Model 3}} = 3.8206$ (1%)   $C_{\text{Model 3}} = 3.8219$ (5%)

$C_{\text{Model 4}} = 3.6786$ (MGCG)   $C_{\text{Model 4}} = 3.6782$ (0.5%)

$C_{\text{Model 4}} = 3.6777$ (1%)   $C_{\text{Model 4}} = 3.6781$ (5%)

Figure 26 Optimization results of the simply supported beam at different scales.

### 4.2.3 3D scenarios with different loading schemes

In this subsection, different loading schemes (as shown in Figure 27) are loaded on the cantilever beam which is fully constrained at the left end. All concentrated forces are axial forces or the resultant force of axial forces. The distributed force is discrete as concentrated forces loading on the nodes. Design domains are all discretized into FE models with hexahedral elements of size 1×1×1. The activation parameter $N_{\text{on}}$ is 20, $\varepsilon_{\text{tol}}$ is 1%, $Ns = 2$, and $Nm = 2$. The maximum CG iterations for MGCG are 50, the convergence tolerance is set to $10^{-6}$, and the grid level is 3.

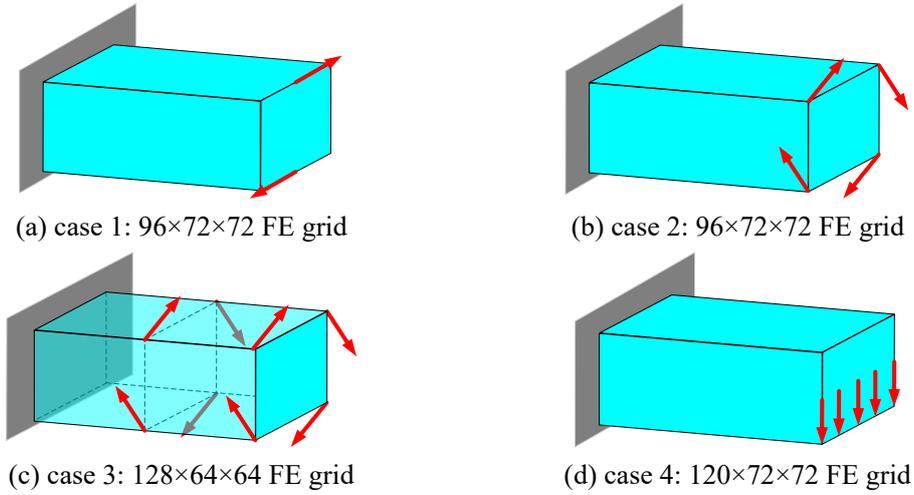

(a) case 1: 96×72×72 FE grid    (b) case 2: 96×72×72 FE grid

(c) case 3: 128×64×64 FE grid    (d) case 4: 120×72×72 FE grid

Figure 27 Geometrical settings for 3D scenarios with different loading schemes.

Figure 28 illustrates the cumulative computational costs and speedup of the two methods. Compared with other cases, the cumulative computational cost of Case 4 is significantly larger due to the excessively large problem scale. Considered in terms of problem scales, the FE grids of cases 1 and 2 are the same, but the difference in speedup also occurs, and the examples tested above indicate that MGCG evaluations are not sensitive to problem scales, so excluding problem scale is not the main reason for the difference in the speedup of the current test. Figure 29 shows that the different load schemes resulted in significant differences in the MGCG evaluations, although the problem size was at the same level. This suggests that the complexity of the loading scheme may affect the fidelity period of AARMR to a certain extent, which in turn affects speedup. Cumulative MGCG evaluations increase as the loading nodes increase, that is, the shorter fidelity period of AARMR. Calling MGCG at high frequency and speedup weakens the speedup. It should be emphasized that the average CG iterations of both AARMR calling MGCG and continuous MGCG reached the limit of 50.

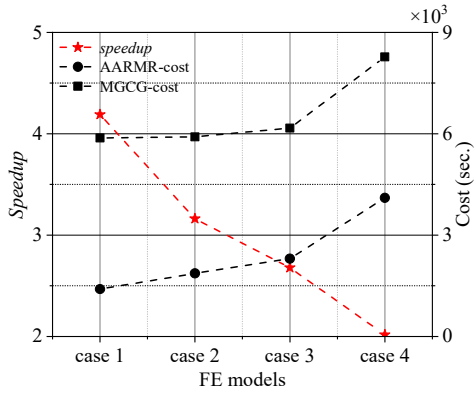 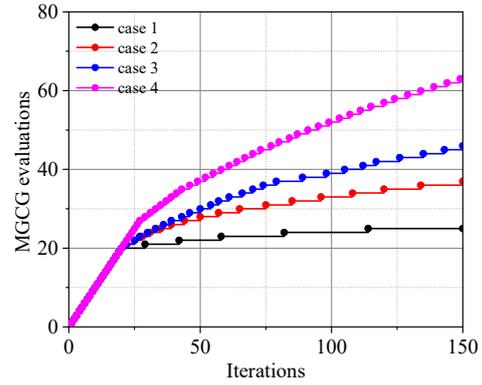

Figure 28 Speedup of different loading schemes.

Figure 29 Cumulative MGCG evaluations for different loading schemes.

Although the loading schemes significantly influence the speedup, the AARMR still achieved a speedup of 2.02 when handling the worst-performing distributed loading scheme (case 4). This indicates that AARMR can significantly improve the efficiency of common load schemes. The accuracy of the optimization results is also acceptable, the maximum difference is only -0.049%. The overall and sectional views of structural topologies shown in Figure 30 have almost no visible difference between the two methods. Furthermore, the consistent symmetry between the loading schemes and the structural topologies indirectly proves the ability of AARMR to capture boundary conditions.

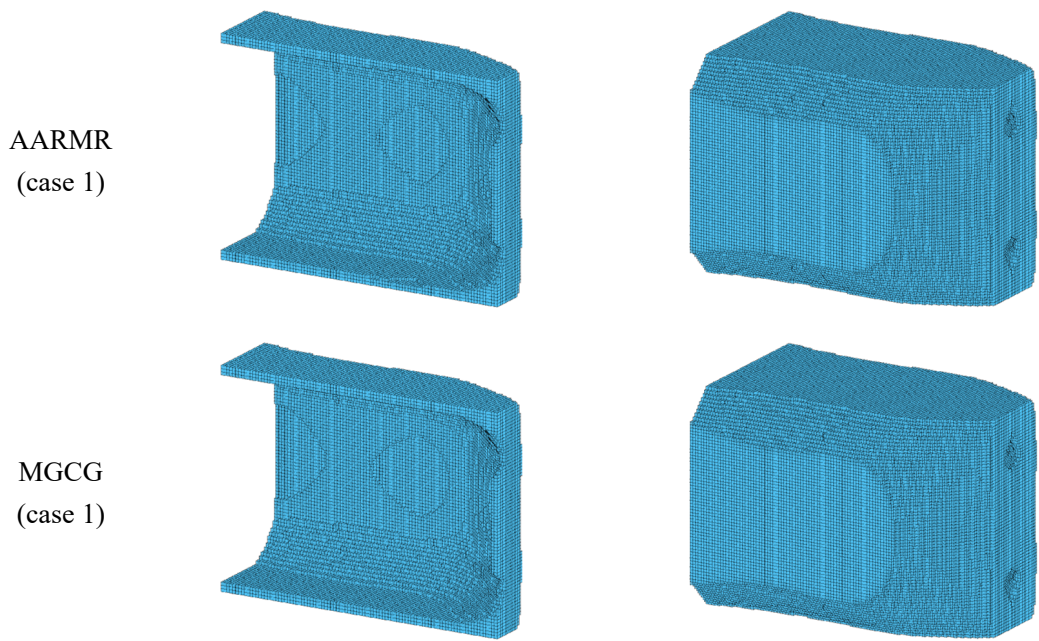

AARMR (case 1)

MGCG (case 1)

| AARMR (case 2) | 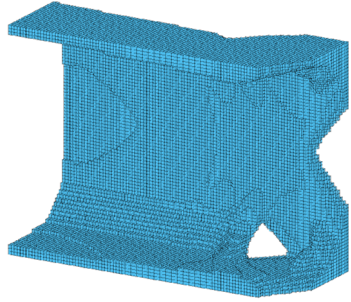 | 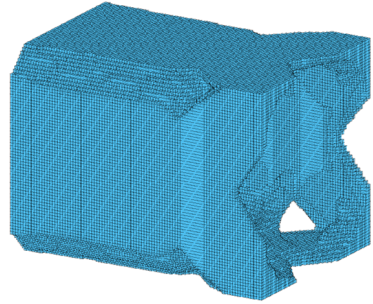 |

| MGCG (case 2) | 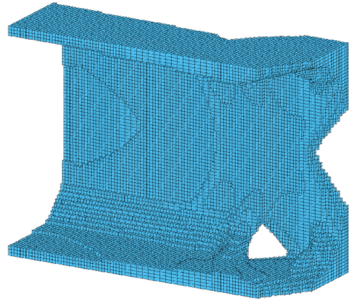 | 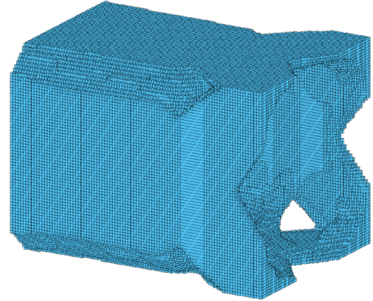 |

| AARMR (case 3) | 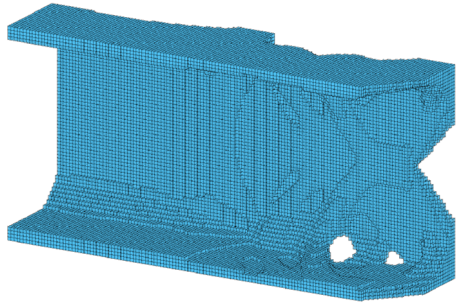 | 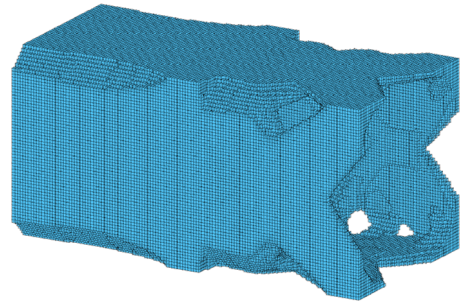 |

| MGCG (case 3) | 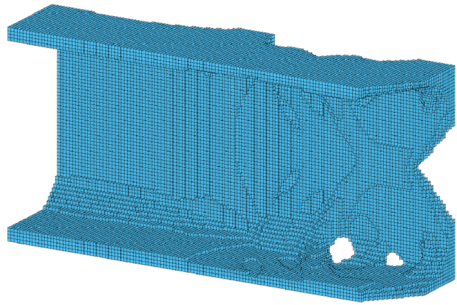 | 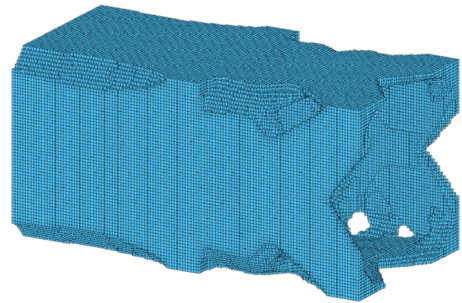 |

| AARMR (case 4) | 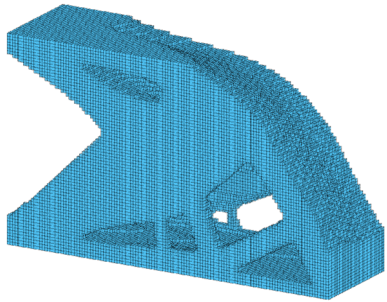 | 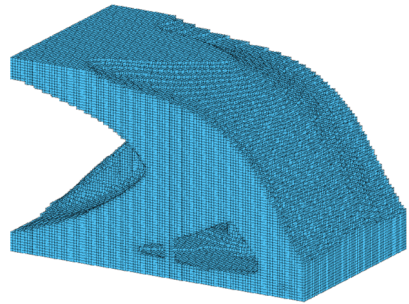 |

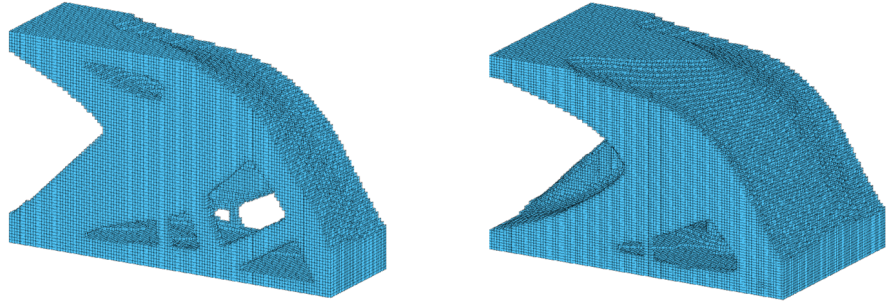

MGCG
(case 4)

Figure 30 Overall and sectional views of optimized structural topologies, the relative differences in end-compliance obtained by AARMR and MGCG, -0.049% for case 1, -0.007% for case 2, -0.036% for case 3 and 0.010% for case 4.

## 4.3 An attempt at nonself-adjoint problems

The main purpose of this paper is to solve the problem of minimizing end-compliance. To expand the application field of AARMR, the nonself-adjoint problem is attempted to be solved. Consider now a typical compliant mechanism design problem (shown in Figure 31). The input end (point A) is subjected to a horizontal static force load $f_{in}$. The objective is to maximize the displacement of the output end (point B). Points A and B are located on the axis of symmetry. Only half of the structure is considered because of symmetry, and the model is discretized into FE models of different scales with linear quadrilateral plane-stress elements of size 1×1. Linear springs simulate the structural stiffness of the input and output ends ($k_{in} = 1$, $k_{out} = 0.1$). The maximum number of design cycles is set to 200, the volume fraction is set to 0.3, the activation parameter $N_{on}$ is 20, $Ns = 2$, and $Nm = 2$. The maximum CG iterations for MGCG are 200, the convergence tolerance is set to $10^{-6}$, and the grid level is 3.

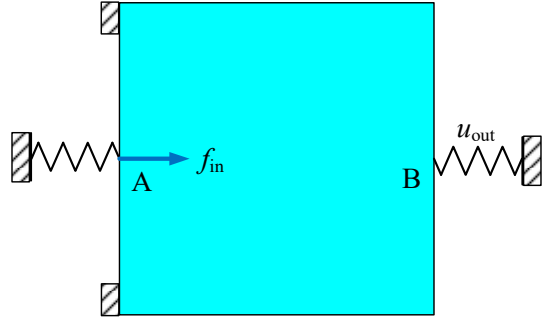

Figure 31 Geometrical settings for 2D displacement–inverter.

For more information about compliant mechanism design, please refer to literature [27]. Approximate sensitivity is employed in this paper, and the optimization results are shown in Table 1 and Figure 32. In this example, the residual criterion is set to 0.01% and 0.1%, and AARMR can still show good acceleration performance, the maximum speedup can reach 1.472. First, the average CG iterations of both methods are at the same level, and even less is required for AARMR. This feature is beneficial to the extended application of AARMR in nonself-adjoint problems. Second, even if the residual standard is strictly required to 0.01%, AARMR can still maintain its advantage over MGCG. Finally, similar to the 2D compliance minimization problems, the speedup appears to be proportional to the problem scales. No matter in terms of the objectives or the topology of optimized structures, AARMR not only shows excellent acceleration performance in the maximizing output displacement, but also the optimization results are close to MGCG and even get a better local optimum. Furthermore, the topology of optimized structure maintains a high consistency regardless of changes in residual criterions or changes in problem scales. This not only shows the grid independence of AARMR, but also implies the stability of the AARMR method.

Table 1 Performance comparison of MGCG and AARMR (L-grid represents low-resolution 320×160 FE grid and H-grid represents high-resolution 640×320 FE grid).

|  | criterion | average CG iterations | | *speedup* | | *diff*. (%) | |
| --- | --- | --- | --- | --- | --- | --- | --- |
|  |  | L-grid | H-grid | L-grid | H-grid | L-grid | H-grid |
| MGCG | —— | 16.187 | 15.373 | —— | —— | —— | —— |
| AARMR | 0.01% | 15.523 | 14.554 | 1.195 | 1.283 | 0.0005 | -0.0030 |
|  | 0.1% | 13.875 | 13.316 | 1.446 | 1.472 | -0.0116 | -0.0544 |

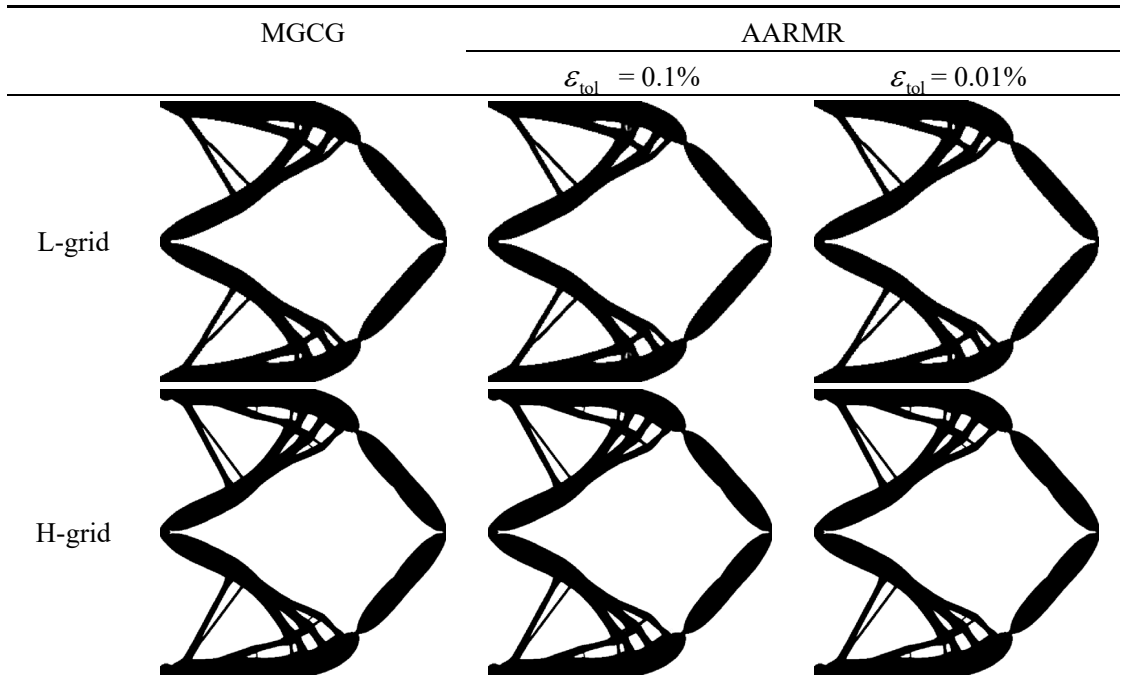

Figure 32 Comparison of results of displacement-inverter problem.

## 5. Conclusion

This study proposed an efficient method called the adaptive auxiliary reduced model reanalysis (AARMR) method, which avoids complex matrix factorization operations and alleviates huge memory requirements. The proposed method not only improves the efficiency of evaluation but can also handle large-scale problems that are difficult to solve by direct solvers on PC-level devices. Simulation results show that the proposed method is more efficient than MGCG.

Several classic numerical examples are tested for comparison with MGCG, and the results prove the high efficiency of AARMR, which is particularly prominent in 3D scenarios. First, AARMR is not sensitive to the basis vector terms of the projection auxiliary reduced model and combined approximation reduced model. Second, the speedup of the test in the 3D scenarios is almost double that of the 2D scenarios, and the maximum speedup can reach 4.85. Surprisingly, the problem scales positively affect on speedup in the 2D scenarios, while they have almost no effect on speedup in the 3D scenarios. This reflects the potential of the AARMR method to significantly save computational costs when solving large-scale problems. Finally, AARMR still shows a great acceleration effect on compliant mechanism design problems (maximizing the displacement of the output end), which shows the extensibility of the method. It is important to emphasize that the AARMR method is not completely against other algorithms, and MGCG is used to update the projection auxiliary reduced model in this paper.

## Declaration of Competing Interest

The authors declare that they have no known competing financial interests or personal relationships that could have appeared to influence the work reported in this paper.

## Acknowledgements

This work has been supported by the Program of National Natural Science Foundation of China under the Grant Numbers 11572120 and 51621004.

## Data Availability Statements

The raw/processed data required to reproduce these findings cannot be shared at this time as the data also forms part of an ongoing study.